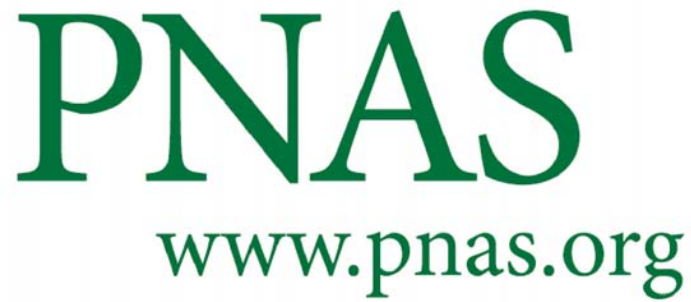

**Main Manuscript for**

Universal Model for Deposition of Entities from Molecular to Nano-sizes on Nanofibers.

Maryam Razavi, Zhao Pan, and Zhongchao Tan.

Mechanical and Mechatronics Engineering, University of Waterloo.

* Zhongchao Tan.

Email: tanz@uwaterloo.ca

Maryam Razavi: https://orcid.org/0000-0002-9040-4994

Zhao Pan: https://orcid.org/0000-0003-1654-3205

Zhongchao Tan: https://orcid.org/0000-0001-7922-8291

**Classification**

Physical Sciences: Applied Physical Sciences

**Author Contributions**

Maryam Razavi: performed research; analyzed data; wrote the paper.

Zhao Pan: analyzed data; wrote the paper.

Zhongchao Tan: designed research; performed research; analyzed data; wrote the paper.

**This PDF file includes:**

    Main Text
    Figures 1 to 6

**Abstract**

Understanding the behavior of entities is critical to the development of new policies and technologies aiming at global health and wellbeing; this paper presents a unified model developed and validated for the deposition of entities, ranging from a few Angstroms to tens of nanometers, on nanofibers. The transport of entities is based on convective diffusion and interfacial interactive diffusion. As the size of entities decreases to intermediate sizes - between gas molecules and nanoparticles - adhesion probability on the surface is less than unity, which is determined by the interfacial interactions and kinetics of adhesion. Dimensionless surface coverage provides a better understanding of adhesion kinetic by considering the effects of initial concentration and time. The model is validated by available experimental data in the literature.

**Significance Statement**

Although larger nanoparticles mostly deposit in the respiratory system, smaller ones can directly enter the brain and cause serious damage. The major fraction of environmental airborne particles is intermediate particles between gas molecules and aerosol nanoparticles. According to the common knowledge of filtration, these particles are fully captured by the available technologies; however, the present study refutes this postulation. Our findings substantially update the current understanding of intermediate nanoparticles and describe the separation of all entities in the air. This understanding will be the foundation of new policies, regulations, and new technologies by providing an overall picture of the entities' deposition process.

**Main Text**

**1. Introduction**

Airborne nanoparticles, as part of particulate matter (PM), have a great impact on human health and the environment (1-4). PM is recognized as the second largest risk factor for non-communicable diseases worldwide. The largest is tobacco smoking, which is also related to inhaled particles (5). larger nanoparticles mostly deposit in the alveolar region of the respiratory system and damage cardiovascular systems (6), while smaller ones impact the nervous system (6-10). The transmission of infectious pathogens through aerosols is another raising health concern (11). Additionally, airborne nanoparticles are responsible for climate change by interfering with thermal radiation (12). Intermediate nanoparticles with the sizes of single-digit nanometers, which bridge gas molecules and aerosol particles, play a critical role in both health and the environment because of their abundance (13).

Intermediate nanoparticles are generated in numerous processes such as combustion and condensation. Measurements show that 20–54% of the ambient aerosol is sub-3 nm nanoparticles (14, 15). These concentrations are higher for those produced in high-temperature combustion (16). This means that intermediate nanoparticle emissions will become a growing challenge because high-temperature combustion is an emerging combustion technology. Intermediate metal nanoparticles are also generated from the melting of engine fragments (17) and sliding contacts (18). For example, up to 80% of the nanoparticles by number emitted from brake friction pairs are in the size range of 1.3 – 4.4 nm (18). In addition, secondary intermediate nanoparticles are formed by condensation of volatile organic compounds (VOCs) (14). Furthermore, intermediate nanoparticles remain in the size that they are generated (12).

Regardless of their great impact on health and the environment, no standard and regulation are available for nanoparticles. However, clean air policies were established to reduce the primary PM2.5 emissions, secondary aerosols, and a variety of gasses (19, 20). In addition, the Institute for Health Metrics and Evaluation (IHME) and the World Health Organization have been assessing annually the global impact of ambient air pollution on health since 2012 (19). Nonetheless, a clear understanding of the behavior of intermediate nanoparticles is essential to knowledge-based policy-making and technology development for air cleaning to further alleviate global environmental and health issues.

Despite the decades of theoretical research in the areas of filtration and adsorption, a holistic framework is still missing for the whole range of entities: there is a gap between the theories of adsorption and filtration. The separation of nanoparticles from the air is typically achieved by particle transport. According to classical air filtration theories, the filtration efficiency of intermediate nanoparticles would approach 100% because of Brownian diffusion; however, these theoretical analyses based on convective diffusion have been validated for a broad range of nanofibers and particle sizes. Some researchers attempted to investigate the effects of interfacial interaction on the transport of nanoparticles to the nanofiber surface (21, 22). More importantly, the assumption of 100% adhesion probability becomes uncertain for intermediate nanoparticles colliding on nanofibers. This lack of fundamental knowledge of adhesion was also pointed out by review studies (23-27).

Admittedly, there are models for the adhesion of intermediate nanoparticles on nanofibers based on interfacial forces (28-31). Dahneke, for example, proposes that intermediate nanoparticles are bound to the surface by molecular attraction (28). Additionally, Hiller and Loeffler state that the filtration efficiency of intermediate nanoparticles depends on the interaction energy between the particles and the fibers(29). Wang and his team (29-30) proposed the thermal rebound theory for sub-10 nm particles. Later, Mouret et al. (32) proposed that thermal rebound might happen for sub-1 nm particles. These earlier works, and many others, remain hypotheses.

To understand the particle deposition based on interfacial forces, we need to revisit the basics of van der Waals force, which may bridge the gap between molecules and normal nanoparticles. The van der Waals is considered as the main factor in the adhesion of sub-micro systems (33, 34) and the adsorption of molecules onto surfaces (34-40). Similarly, interfacial interaction should play an important role in the adhesion of intermediate nanoparticles, which are expected to behave like large molecules. Non-perfect adhesion has been considered in adsorption models, where only a fraction of striking molecules are adsorbed because of the weakened interfacial interaction at the intermediate size (35, 36). For example, Chernyakov mentioned that desorption should result in decreased filtration efficiency of intermediate nanoparticles. In his model, the residence time of particles in the bond state is considered, where the efficiency reduction takes place based on desorption for long times of filter operation (37); however, experimental results (38) show a reduction of efficiency even for short times of operation.

In addition to the size, operating time and particle concentration may also affect the deposition of intermediate nanoparticles on nanofibers because both factors play an important role in adsorption kinetics. In the filtration also concentration dependency of nanoparticle filtration has been observed on fibrous fibers (22, 38, 39).

This paper presents a unified model that unifies all state-of-the-art theories, including gas adsorption, intermediate nanoparticle filtration, and normal aerosol filtration. The new model is developed by considering the steady deposition of entities, kinetic behavior of adhesion, and interfacial interaction. This universal model is also validated using experimental data in the literature.

## 2 Results and Discussion

This model explained the steady-state transport of entities based on diffusion by considering two flow fields around the nanofiber. The transport behavior of nanoparticles based on airflow and interaction force of Van der Waals is validated by the experimental data available in the literature for nanoparticles (Section 2.1). From Equation (9), all nanoparticles adhere to the surface;

therefore, single fiber efficiency is reduced to transport mechanisms ($\eta_0 = \eta_c$). The properties and operating conditions of experimental results from the literature are listed in the supplementary information (S6).

The adhesion probability of nanoparticles drops below unity when the size of nanoparticles decreases to the intermediate size and under. The adhesion probability depends on size, time, and concentration, and the surface coverage function describes the related kinetic behavior of deposition. Since the measured efficiency depends only on surface coverage (in terms of time and concentration), the surface coverage function (Equation 10) is compared with the transformed surface coverage (Section 2.2):

$$f_{exp}(y) = \frac{\eta_{exp}}{\eta_c S_0 \frac{4\alpha L}{\pi d_f (1-\alpha)}}, \qquad (1)$$

where $\alpha$ is the porosity of filter, $L$ is the filter thickness, and $d_f$ is the fiber diameter. Then, the filtration efficiency of intermediate nanoparticles is validated by experiments in Section 2.3. Finally, the general model is explained and compared with the common knowledge in the filtration and adsorption (Section 2.4).

**2.1 Validation of the transport model**

Peclet number is used in classic filtration theories for the calculation of single fiber efficiency that is attributed to diffusion. According to the previous empirical and theoretical models in the filtration field, $\eta_0 \sim Pe^{-m}$, where $m = -2/3$ (40-43) or $m = -0.43$ (44). They are plotted in Figure 1a against experimental data (44-51). The experimental data do not necessarily scale with a certain power of Peclet number. In our model, therefore, parameter $R$ (the ratio of particle to fiber diameter) also plays a role in the diffusion of entities. Figure 1b shows the single fiber efficiency derived from the same experimental data versus parameter $R$. This new presentation reveals that the deposition of nanoparticles is separated into three regimes defined by $R$.

Figure 2 shows the role of $Pe$ number and validation of the model for each regime. When $R$ is small $[R < O(10^{-4})]$, the dimensionless rate of collision due to the van der Waals dominates the deposition mechanism, and the corresponding single fiber efficiency is proportional to the non-dimensional particle collision rate: $\eta_0 \propto N'_{t-vdw} \sim Pe^{-1}$. Figure 2a shows that the experimental data of the first regime collapse on the solid line, which represents a $Pe^{-1}$ scaling law. The dimensionless rate of collision based on van der Waals interaction (Equation 6) is validated in Figure 2d. For high $Pe$ and $R$, however, the experimental data deviate from the model (see supplementary information, S7). Equation (7) indicates that the dimensionless rate of collision based on van der Waals becomes negligible by increasing the particle size beyond a critical value. These data are in the boundary of first and second regimes.

By increasing R values, $O(10^{-4}) < R < O(10^{-2})$, the dimensionless rate of collision based on airflow dominates the deposition mechanism, and the corresponding single fiber efficiency is proportional to the non-dimensional particle collision rate: $\eta_0 \propto N'_{t-vdw} \sim Pe^{-1}$. As seen in Figure 2b, the experimental data of the second regime collapse on the solid line, which represents a scaling law of $Pe^{-1/2}$. The dimensionless rate of collision based on air flow from Equation (6) is validated in Figure 2e. The deviation from linear behavior for smaller $Pe$ is due to the drop of adhesion probability.

Further increasing parameter $R$, $O(10^{-2}) < R$, the interception mechanism also becomes important in addition to diffusion. The efficiency due to the interception mechanism is proportional to $\frac{R^2}{1+R}$ (52). In Figure 2c, the experimental data for the third regime collapse on the solid line, which also represents the scaling law of $Pe^{-1/2}$. The dimensionless rate of collision based on air flow from Equation (6) is validated in Figure 2f. The deviation from linear behavior is likely due to increasing efficiency by the interception mechanism, and it is out of the scope of this study.

**2.2 Validation of the model for surface coverage**

Figure 3 compares the model (Equation 10) with the surface coverage function transformed from experimental data (53, 54) from Equation (1). Surface coverage of the experimental data, x-axis, is calculated using the time and concentration and other parameters reported in the references for a single size entity. See Equation (S-76), where $a = 1$, $b = 0.25$, and the size of toluene is assumed 6.7 Å. These experimental data are from physisorption of gas molecules on the nanofibers. Using Equation (1), the nanofiber properties were used to correlate the data for a single fiber. In the adsorption field, specific surface area ($A_{bet}$) is one of the most important adsorbent properties for understanding kinetic behavior (53), while in the filtration α, L, and $d_f$ are the common reported filter's properties. Thus, $\frac{4\alpha L}{\pi d_f(1-\alpha)} = \frac{A_{bet} m}{\pi A_f(1-\alpha)}$ is used in the correlation of the single fiber and total efficiency (see supplementary information, S8), where $m$ is the mass of filter, and $A_f$ is the filter surface perpendicular to the flow. Since the new coefficient is a very large number, the normalized $f_{exp}(y)$ is shown in Figure 3. The solid line shows the model result for $f(y) = \left(1 + \frac{y}{1-y} PR\right)^{-1}$, where $PR = \frac{S_0 P_b'}{P_a}$ is the probability ratio. Assuming monolayer deposition, $PR$ is estimated as 0.636. (see supplementary information, S4, for details). The dashed line shows the model, with lower values of $PR = 0.370$. Both the model and experimental data show that the surface coverage function, $f(y)$, which is proportional to the adhesion probability, drops with decreasing surface coverage. The surface coverage gradient of the experimental data is low (y=0 to 0.5) for low surface coverages. At a high surface coverage, however, the gradient drops sharply (y=0.5 to 1).

Several factors may contribute to the discrepancy between $f_{exp}(y)$ and $f(y)$. First, the model assumes monolayer deposition, but real deposition may occur on multiple layers. Although a surface coverage function based on multilayer assumption is derived in this study (see Equation S-66), a simple surface coverage function for monolayer is employed due to the difficulty of estimation on the energies in each state. Further research on the surface coverage function is needed to calculate a more robust adhesion probability. Second, the energy difference between adhesion and detached states are estimated by interfacial interaction, rather than the exact values for the energy of each state (see supplementary information, S4). Additionally, for simplicity, the Hamaker model is used for interfacial energies to calculate the probability of different states; a better estimation of the interfacial interaction can improve the present model. Third, the discrepancy may also be attributed to the properties of filter media which in some cases, a range was reported instead of a single value for the filter properties (see Table S3). More experimental data with the measured filter properties is needed for the validation of the model.

**2.3 Validation of the model for intermediate nanoparticles**

Supplementary information (S10) shows the validations for intermediate nanoparticles with all experimental data we can find in literature, and one of them is explained here. Figure 4 compares the measured efficiencies (38) with the calculations for intermediate nanoparticles in the size range of 0.8-3.3 nm (i.e., $Pe = 0.003 - 0.05$). The dashed line is for time zero and the solid line, at 10 min. Constants $a = 9 \times 10^{-8}$ and $b = 1.6 \times 10^8$ are used for all six cases. The scattered markers are for the experimental data. The operation time was not reported in the literature (43). We calculated the error bars owing to the systematic error associate with the limit of detection (LOD) (detailed calculation is available in the supplementary information, S9).

The model shows that, at time zero, the filtration efficiency drops with decreasing particle size of the intermediate nanoparticles. This is likely due to the lower initial adhesion probability. However, the experimental data show shaper reductions for $Pe$ numbers less than 0.02 (sub-2 nm particles).

Over time, the total efficiency drops while more and more particles deposit on the nanofibers. The comparison shows that experimental data agreed with the model results after a short period (~10 min), which is consistent with all six cases. Therefore, it can be inferred that the effects of concentration on the deposition of intermediate nanoparticles become influential over time. Nonetheless, this behavior might not be observed for lower concentration distribution than these six cases since surface coverage effect is negligible at lower concentrations and a short time.

**2.4 Validation of the model for all size ranges**

Figure 5 shows the general model results for the deposition of entities on the nanofiber and compared with the experimental data for both filtration and adsorption. Figure 5b presents the total efficiency as a function of entity size and time, for two concentrations. The high concentration is approximately 10 times that of the low concentration. The WOx entities are in the size range of 0.4 - 110 nm, and the mean diameter of nanofiber is 191 nm. The general model becomes an adsorption model (Figure 5a) and a filtration model (Figure 5c) when the entity size approaches the molecular size and larger nanoparticles, respectively. The unified model is further explained separately for molecular, nano, and intermediate sizes as follows.

For the molecular sizes, the model shows that the total efficiency is time and concentration dependant. It can predict the breakthrough behavior in adsorption kinetics. For example, Figure 5a shows the efficiency for $d_p = 0.4\ nm$ for a high initial concentration (solid black line) and low initial concentration (dash red line). Besides time, the initial concentration also affects surface coverage and efficiency. The efficiencies are lower for the high initial concentration because of relatively higher surface coverage. These results are consistent with the breakthrough curves in adsorption kinetics. The breakthrough takes place faster at higher concentrations and the ratio of downstream concentration to the upstream concentration (or penetration) is lower than the values for low concentrations during the time of operation (55).

Figure 5d shows the total efficiency of a function of surface coverage by considering both time and concentration. The effect of surface coverage is negligible when the concentration is low. Figure 5g shows the measured total adsorption efficiency as a function of their calculated surface coverage, which is discussed in Section 2.2. All these prove that the unified model applies to adsorption. The efficiency of large nanoparticles is generally believed to vary with particle size, but it is independent of concentration or time. Figure 5c shows that our unified model concurs this classic theory, where particles are diffused to the nanofiber. The diffusion of particles to the nanofiber surface can be determined based on the airflow or van der Waals interaction with the surface. For a fresh surface, a reduction in efficiency is observed while the particle size approaches molecular size. The efficiency drops from 1 nm. This can be explained as follows. As the size of nanoparticles drops to the intermediate size range, the adhesive interaction energy binding particles to the fiber's surface continually drops, while the average energy of particles remains constant. As a result, the intermediate nanoparticles begin to detach from the surface more easily than larger ones. For larger particle sizes, adhesion probability approaches to unity, therefore diffusion only in the boundary layer determines the filtration, which does not vary with the time and concentration. Therefore, the efficiency of nanoparticles shown in Figure 5c is stable, whereas Figure 5b reveals the efficiency drop further over time for intermediate nanoparticles.

The efficiency drop in the intermediate size range depends on the properties of particles and nanofibers. However, the efficiency of intermediate nanoparticles might not drop when particles and fibers have great Hamaker constants, which increase adhesion probability for particles to detach from the surface. In this model, the Hamaker model was used for adhesive interaction energy between bodies, which has simplification assumptions to express the interaction of particles. Discreetness of transient nanoparticles, non-additivity, and retardation are ignored in this model (56, 57). A more accurate definition for the interaction of transient nanoparticles with a surface should be developed in the future to improve this model.

Figure 5d shows the total efficiency as a function of $Pe$. The collision efficiency of nanoparticles based on either airflow (black line) or van der Waals interaction (purple line) is a function of $Pe$. Figure 5i shows experimental results for total filtration efficiency as a function of $Pe$, which is discussed in Section 2.1.

Our model shows that the deposition of intermediate nanoparticles depends on size, time, and concentration. Figure 5c shows that the total efficiency of intermediate nanoparticles drops because the adhesion probability is lower than unity in this size range at time zero. Figure 5b depicts their total efficiencies with different particle sizes, times, and concentrations. At time zero, both efficiencies for low concentration (dashed red line) and high concentration (solid black line) are entangled. As time goes on, the effect of concentration on the deposition becomes important.

Similar to the adhesion of molecules, the adhesion probability of intermediate nanoparticles is lower when they collide on an occupied site than that on a fresh surface. Thus, more surface coverage causes lower adhesion probability for the intermediate nanoparticles. This phenomenon is adopted in the unified model: adhesion probability is determined by interfacial interaction energy, which is the barrier for detachment. Moreover, if a particle collides on an occupied site, the separation between the collided particle and surface increases: the adhesive interaction energy decreases with the power of 6 ($E_{int} = -C/r^6$). All in all, It can be concluded that, regardless of monolayer or multilayer deposition, the adhesion probability and total efficiency of deposition decrease with increased surface coverage. Nevertheless, this paper reports the work based on monolayer deposition for simplicity.

Figure 5e shows the total efficiency calculated for the deposition of entities on nanofiber in terms of $Pe$ and surface coverage. The deposition of intermediate nanoparticles depends on both $Pe$ and surface coverage. With zero surface coverage, the efficiency drop decreases with size. At a relatively higher surface coverage, however, the efficiency first drops steeply then slightly increases at a very low $Pe$. The reason for this minor increase at low $Pe$ number is the reduced surface coverage for smaller particles.

As a final validation, Figure 5h summarizes all the measured total efficiencies for nanoparticles, intermediate nanoparticles, and molecules as a function of $Pe$ and surface coverage. Overall, they agree with the model. Note that these experimental data are also analyzed in Section 2.3.

**3 Model Development**
Figure 6 illustrates the unified model. The total filtration efficiency is determined by the correlation of single fiber and filter media (left) based on mass balance. Such an approach allows the model to be used for entities in a wide size range. The model assumes a steady transport of entities to the fiber's surface (top-right) followed by a kinetic process of adhesion (bottom-right). The transport of entities is based on diffusion: convective diffusion and interfacial interactive diffusion. They are determined using boundary layer analysis around a single fiber (see Section 3.1). It is assumed that a fraction of entities that collide on the fiber surface adhere to it. The particle collection efficiency of a single fiber, $\eta_0$, is a product of collision efficiency $\eta_c$ and adhesion probability $\eta_{ad}$:

$$\eta_0 = \eta_{ad}\eta_c. \qquad (2)$$

Adhesion probability is modeled based on the particle-surface interfacial interaction, the time of operation, and the initial concentration of entities. The kinetic adhesion of entities is explained in Section 3.2.

The theoretical analysis in this paper is based on these assumptions: isothermal condition, uniform temperature, homogenous fiber surface, and equal surface energy. In addition, interaction among entities is ignored.

## 3.1 Steady transport of entities onto the single fiber

Consider a particle-laden flow (with a free-stream particle concentration, $n_0$) passing a cylindrical fiber. Particles colliding the fiber due to the diffusion would adhere to the surface, resulting in reduced particle concentration in the air. Figure 6 the concentration boundary layer (CBL) developed around the fiber. The particle concentration on the outer edge of the CBL equals that in the free stream, and the concentration gradient is negligible. The concentration gradient in the CBL drives particles towards the fiber surface. Outside the CBL, there is no radial flux and diffusion. The particle concentration diffusion process can be modeled as (58):

$$u_r \frac{\partial n}{\partial r} + \frac{u_\theta}{r} \frac{\partial n}{\partial \theta} = D \left( \frac{\partial^2 n}{\partial r^2} + \frac{1}{r} \frac{\partial n}{\partial r} + \frac{\partial^2 n}{r^2 \partial \theta^2} \right), \quad (3)$$

where $n$ is the number concentration; $u_r$ and $u_\theta$ are radial and tangential velocity components, respectively; $D$ is the diffusion coefficient. The aforementioned CBL gives $\frac{\partial n}{\partial r} = 0$, $n \to n_0(\theta)$ at $r = \rho(\theta)$ and $n = n_r \neq 0$ when $r = R_f$. By employing stream function over the fiber instead of velocity components ($u_r = \frac{1}{r} \frac{\partial \psi}{\partial \theta}$, $u_\theta = -\frac{\partial \psi}{\partial r}$) and further simplifications, Equation (3) is reduced:

$$-\frac{\partial}{\partial \theta} \int_{n_r}^{n_o} \psi \, dn = D R_f \left( \frac{\partial n}{\partial r} \right)_{r=R_f}. \quad (4)$$

Details of this transformation are available in the supplementary information (S1). The rate of particles entering the boundary layer at each $\theta$ is $N = \int_{n_r}^{n_o} \psi \, dn$. The total rate of particles colliding on the surface over the fiber surface is obtained by integration:

$$N_t = 2 \int_0^\pi \frac{\partial N}{\partial \theta} \, d\theta, \quad (5)$$

which is attributed to two stream functions. The rate of particles to the fiber due to each of them is determined separately because each has its boundary layer thickness. The stream function of airflow over the fiber is approximated using the Kuwabara flow field (59), and the total dimensionless rate of particles colliding on the fiber's surface based on this air stream is calculated as:

$$N'_{t-f} = 4 \left( \frac{Pe \, Ku}{1 - \alpha} \right)^{-1/2}, \quad (6)$$

where $Pe = u R_f / D$ is the Peclet number, $\alpha$ is the solidity of filter, and $Ku$ is the Kuwabara hydrodynamic factor. Derivation of Equation (6) is provided in the supplementary information (S2). Moreover, the total dimensionless rate of particle collision on fiber due to the van der Waals force can be described as a function of Pe and $R = d_p/d_f$:

$$N'_{t-vdw} = \frac{c_1 \pi^{\left(-\frac{6k_bT}{A}(R+1)\right)}}{\frac{6k_bT}{A}(R+1) + 1} + \frac{2\pi Pe^{-1}}{\frac{3k_bT}{A}R(R+1) + \frac{R}{2}}, \quad (7)$$

where $k_b$ is the Boltzmann's constant, $T$ is the temperature, $A$ is the Hamaker constant (60), and $c_1 = 1.5 \times 10^4$ is a constant coefficient (see supplementary information, S3, for more details). The collision efficiency of particles that diffused to fiber by considering either airflow (Equation (6)) or van der Waals force (Equation (7)) is calculated by:

$$\eta_c = 1 - \left(1 - C'_d N'_{t-vdw}\right)\left(1 - C_d N'_{t-f}\right), \tag{8}$$

where $C'_d$ and $C_d$ are correction factors.

**3.2 Kinetic adhesion between entities and a fiber**

The adhesion probability of entities is given as:

$$\eta_{ad} = S_0 f(y), \tag{9}$$

where $S_0$ is the initial adhesion probability and $f(y)$ is the function of surface coverage, y. See supplementary information for more details, S4. $S_0$ is defined as the ratio of the number of adhered particles to the number of impinging particles at zero coverage:

$$S_0 = \frac{P_a}{P_a + P_b}, \tag{10}$$

where $P_a$ and $P_b$ are the probabilities of a particle adheres onto and detaches from an empty site on a fiber, respectively. Assuming monolayer adsorption, which means that the probability of adhesion to an occupied site is zero, the surface coverage function is (61):

$$f(y) = \left(1 + \frac{y}{1-y} \frac{S_0 P'_b}{P_a}\right)^{-1}, \tag{11}$$

where $P_b'$ is the probability of detaching from an occupied site (see supplementary information for more details, S4). The probabilities in Equations (10) and (11) are calculated using the partition function (see supplementary information for more details, S4). The surface coverage can also be determined for a kinetic deposition process. The surface coverage for polydisperse particles is

$$y = b \left( \int_0^{+\infty} n_0^{\frac{2}{3}} \eta_{0(t=0)} \pi (d_p)^2 \, dd_p \right) (1 - e^{-kt}), \tag{12}$$

where $b$ is a constant that reinforces the saturated surface coverage when $t \to \infty$ and $k$ is the overall effective mass transfer coefficient ($k = a \frac{A_f}{A_{bet}} \cdot \frac{\eta_{0(t=0)} u}{d_f}$). A detailed calculation of $b$ is shown in the supplementary information (S5).

**Acknowledgments**

The authors would like to acknowledge the financial support from the Natural Sciences and Engineering Research Council of Canada (RGPIN-2020-04687) and the GCI Ventures Capital.

**References**

1. Fiore AM, *et al.* (2012) Global air quality and climate. *Chemical Society Reviews* 41(19):6663-6683.
2. Huang R-J, *et al.* (2014) High secondary aerosol contribution to particulate pollution during haze events in China. *Nature* 514(7521):218.
3. Lelieveld J, Evans JS, Fnais M, Giannadaki D, & Pozzer A (2015) The contribution of outdoor air pollution sources to premature mortality on a global scale. *Nature* 525(7569):367.

4. Maynard AD*, et al.* (2006) Safe handling of nanotechnology. *Nature* 444(7117):267.
5. Prüss-Ustün A*, et al.* (2019) Environmental risks and non-communicable diseases. *Bmj* 364.
6. Maher BA*, et al.* (2016) Magnetite pollution nanoparticles in the human brain. *Proceedings of the National Academy of Sciences* 113(39):10797-10801.
7. Calderon-Garciduenas L*, et al.* (2003) DNA damage in nasal and brain tissues of canines exposed to air pollutants is associated with evidence of chronic brain inflammation and neurodegeneration. *Toxicologic pathology* 31(5):524-538.
8. Chen H*, et al.* (2017) Living near major roads and the incidence of dementia, Parkinson's disease, and multiple sclerosis: a population-based cohort study. *The Lancet* 389(10070):718-726.
9. Sunyer J*, et al.* (2015) Association between traffic-related air pollution in schools and cognitive development in primary school children: a prospective cohort study. *PLoS Med* 12(3):e1001792.
10. Peeples L (2020) News Feature: How air pollution threatens brain health. *Proceedings of the National Academy of Sciences*:202008940.
11. Morawska L & Milton DK (2020) It is time to address airborne transmission of COVID-19. *Clinical Infectious Diseases* 7.
12. Kulmala M*, et al.* (2013) Direct observations of atmospheric aerosol nucleation. *Science* 339(6122):943-946.
13. Pope Iii CA*, et al.* (2002) Lung cancer, cardiopulmonary mortality, and long-term exposure to fine particulate air pollution. *Jama* 287(9):1132-1141.
14. Rönkkö T*, et al.* (2017) Traffic is a major source of atmospheric nanocluster aerosol. *Proceedings of the National Academy of Sciences* 114(29):7549-7554.
15. Hietikko R*, et al.* (2018) Diurnal variation of nanocluster aerosol concentrations and emission factors in a street canyon. *Atmospheric environment* 189:98-106.
16. Sgro LA*, et al.* (2008) Measurements of nanoparticles of organic carbon and soot in flames and vehicle exhausts. *Environmental science & technology* 42(3):859-863.
17. Liati A, Pandurangi SS, Boulouchos K, Schreiber D, & Dasilva YAR (2015) Metal nanoparticles in diesel exhaust derived by in-cylinder melting of detached engine fragments. *Atmospheric environment* 101:34-40.
18. Nosko O, Vanhanen J, & Olofsson U (2017) Emission of 1.3–10 nm airborne particles from brake materials. *Aerosol Science and Technology* 51(1):91-96.
19. Anenberg S, Miller J, Henze D, & Minjares R (2019) A global snapshot of the air pollution-related health impacts of transportation sector emissions in 2010 and 2015. *International Council on Clean Transportation: Washington, DC, USA*.
20. Scovronick N*, et al.* (2019) The impact of human health co-benefits on evaluations of global climate policy. *Nature communications* 10(1):1-12.
21. Spielman LA & Goren SL (1970) Capture of small particles by London forces from low-speed liquid flows. *Environmental Science & Technology* 4(2):135-140.
22. Sinha-Ray S, Sinha-Ray S, Yarin AL, & Pourdeyhimi B (2015) Application of solution-blown 20–50 nm nanofibers in filtration of nanoparticles: The efficient van der Waals collectors. *Journal of membrane science* 485:132-150.
23. Biswas P & Wu C-Y (2005) Nanoparticles and the environment. *Journal of the Air & Waste Management Association* 55(6):708-746.

24. Wang C-s & Otani Y (2012) Removal of nanoparticles from gas streams by fibrous filters: a review. *Industrial & Engineering Chemistry Research* 52(1):5-17.
25. Abdolghader P, Brochot C, Haghighat F, & Bahloul A (2018) Airborne Nanoparticle Filtration Performance of Fibrous Media: A review. *Science and Technology for the Built Environment* (just-accepted).
26. Preining O (1998) The physical nature of very, very small particles and its impact on their behaviour. *Journal of Aerosol Science* 29(5-6):481-495.
27. Preining O (2008) The physical nature of very, very small particles and its impact on their behavior. *Developments in Surface Contamination and Cleaning: Fundamentals and Applied Aspects*,  (Elsevier), pp 3-24.
28. Dahneke B (1975) Kinetic theory of the escape of particles from surfaces. *Journal of Colloid and Interface Science* 50(1):89-107.
29. Hiller R & Löffler F (1980) Influence of particle impact and adhesion on the collection efficiency of fibre filters. *German Chemical Engineering* 3:327-332.
30. Wang H-C (1986) Theoretical adhesion efficiency for particles impacting a cylinder at high Reynolds number. *Journal of Aerosol Science* 17(5):827-837.
31. Wang H-C & Kasper G (1991) Filtration efficiency of nanometer-size aerosol particles. *Journal of Aerosol Science* 22(1):31-41.
32. Mouret G, Chazelet S, Thomas D, & Bemer D (2011) Discussion about the thermal rebound of nanoparticles. *Separation and Purification Technology* 78(2):125-131.
33. DelRio FW*, et al.* (2005) The role of van der Waals forces in adhesion of micromachined surfaces. *Nature materials* 4(8):629.
34. Wang S, Wang C, Peng Z, & Chen S (2018) A new technique for nanoparticle transport and its application in a novel nano-sieve. *Scientific reports* 8(1):9682.
35. Do DD (1998) *Adsorption analysis: equilibria and kinetics* (Imperial college press London).
36. Kisliuk P (1957) The sticking probabilities of gases chemisorbed on the surfaces of solids. *Journal of Physics and Chemistry of Solids* 3(1-2):95-101.
37. Chernyakov A (2016) Filtration of nanoaerosols through porous materials with allowance for desorption processes. *Colloid Journal* 78(5):717-721.
38. Givehchi R, Li Q, & Tan Z (2018) Filtration of Sub-3.3 nm Tungsten Oxide Particles Using Nanofibrous Filters. *Materials* 11(8):1277.
39. Ardkapan SR, Johnson MS, Yazdi S, Afshari A, & Bergsøe NC (2014) Filtration efficiency of an electrostatic fibrous filter: Studying filtration dependency on ultrafine particle exposure and composition. *Journal of Aerosol Science* 72:14-20.
40. Kirsch A & Fuchs N (1968) Studies on fibrous aerosol filters—III Diffusional deposition of aerosols in fibrous filters. *Annals of Occupational Hygiene* 11(4):299-304.
41. Lee K & Liu B (1982) Theoretical study of aerosol filtration by fibrous filters. *Aerosol Science and Technology* 1(2):147-161.
42. Payet S, Boulaud D, Madelaine G, & Renoux A (1992) Penetration and pressure drop of a HEPA filter during loading with submicron liquid particles. *Journal of Aerosol Science* 23(7):723-735.
43. Hinds WC (1999) Aerosol Technology: Properties. *Behavior, and Measurement of airborne Particles (2nd*.
44. Wang J, Chen D, & Pui D (2007) Modeling of filtration efficiency of nanoparticles in standard filter media. *Journal of Nanoparticle Research* 9(1):109-115.

45. Heim M, Attoui M, & Kasper G (2010) The efficiency of diffusional particle collection onto wire grids in the mobility equivalent size range of 1.2–8nm. *Journal of Aerosol Science* 41(2):207-222.
46. Shin W, Mulholland G, Kim S, & Pui D (2008) Experimental study of filtration efficiency of nanoparticles below 20nm at elevated temperatures. *Journal of Aerosol Science* 39(6):488-499.
47. Van Gulijk C, Bal E, & Schmidt-Ott A (2009) Experimental evidence of reduced sticking of nanoparticles on a metal grid. *Journal of Aerosol Science* 40(4):362-369.
48. Kim SC, Harrington MS, & Pui DY (2007) Experimental study of nanoparticles penetration through commercial filter media. *Journal of Nanoparticle Research* 9(1):117-125.
49. Rengasamy S, King WP, Eimer BC, & Shaffer RE (2008) Filtration performance of NIOSH-approved N95 and P100 filtering facepiece respirators against 4 to 30 nanometer-size nanoparticles. *Journal of occupational and environmental hygiene* 5(9):556-564.
50. Steffens J & Coury J (2007) Collection efficiency of fiber filters operating on the removal of nano-sized aerosol particles: I—Homogeneous fibers. *Separation and purification technology* 58(1):99-105.
51. Givehchi R, Li Q, & Tan Z (2016) Quality factors of PVA nanofibrous filters for airborne particles in the size range of 10–125 nm. *Fuel* 181:1273-1280.
52. Givehchi R & Tan Z (2014) An overview of airborne nanoparticle filtration and thermal rebound theory. *Aerosol Air Qual Res* 14:45-63.
53. Das D, Gaur V, & Verma N (2004) Removal of volatile organic compound by activated carbon fiber. *Carbon* 42(14):2949-2962.
54. Dwivedi P, Gaur V, Sharma A, & Verma N (2004) Comparative study of removal of volatile organic compounds by cryogenic condensation and adsorption by activated carbon fiber. *Separation and Purification Technology* 39(1-2):23-37.
55. Azizian S (2004) Kinetic models of sorption: a theoretical analysis. *Journal of colloid and Interface Science* 276(1):47-52.
56. Kim H-Y, Sofo JO, Velegol D, Cole MW, & Lucas AA (2007) Van der Waals dispersion forces between dielectric nanoclusters. *Langmuir* 23(4):1735-1740.
57. Israelachvili JN (2011) *Intermolecular and surface forces* (Academic press).
58. Friedlander S (1957) Mass and heat transfer to single spheres and cylinders at low Reynolds numbers. *AIChE journal* 3(1):43-48.
59. Kuwabara S (1959) The forces experienced by randomly distributed parallel circular cylinders or spheres in a viscous flow at small Reynolds numbers. *Journal of the physical society of Japan* 14(4):527-532.
60. Lifshitz E (1955) The theory of molecular attractive forces between solid bodies. *J. Exp. Theor. Phys. USSR* 29:83-94.
61. Kirsh V (2000) The effect of van der Waals' forces on aerosol filtration with fibrous filters. *Colloid Journal* 62(6):714-720.

**Figures and Tables**

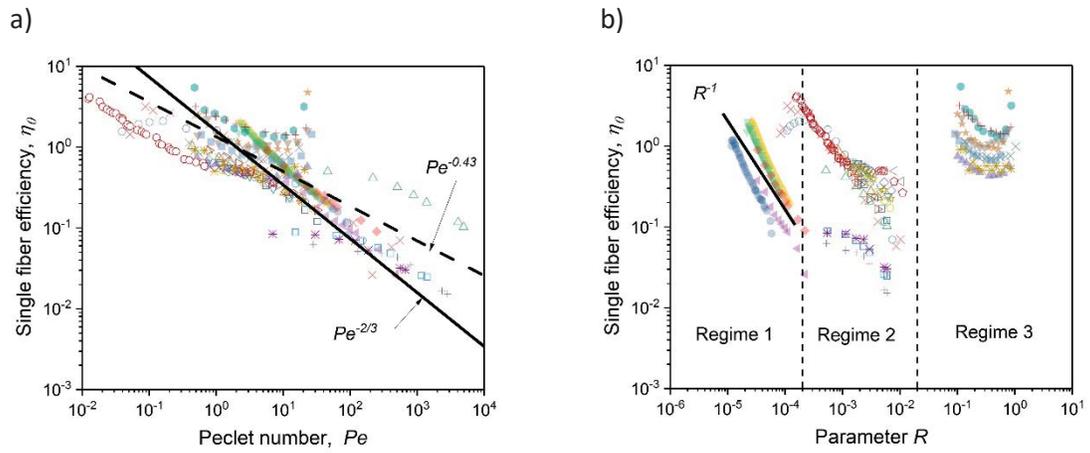

**Figure 1.** The experimental data versus (a) Pe and (b) R

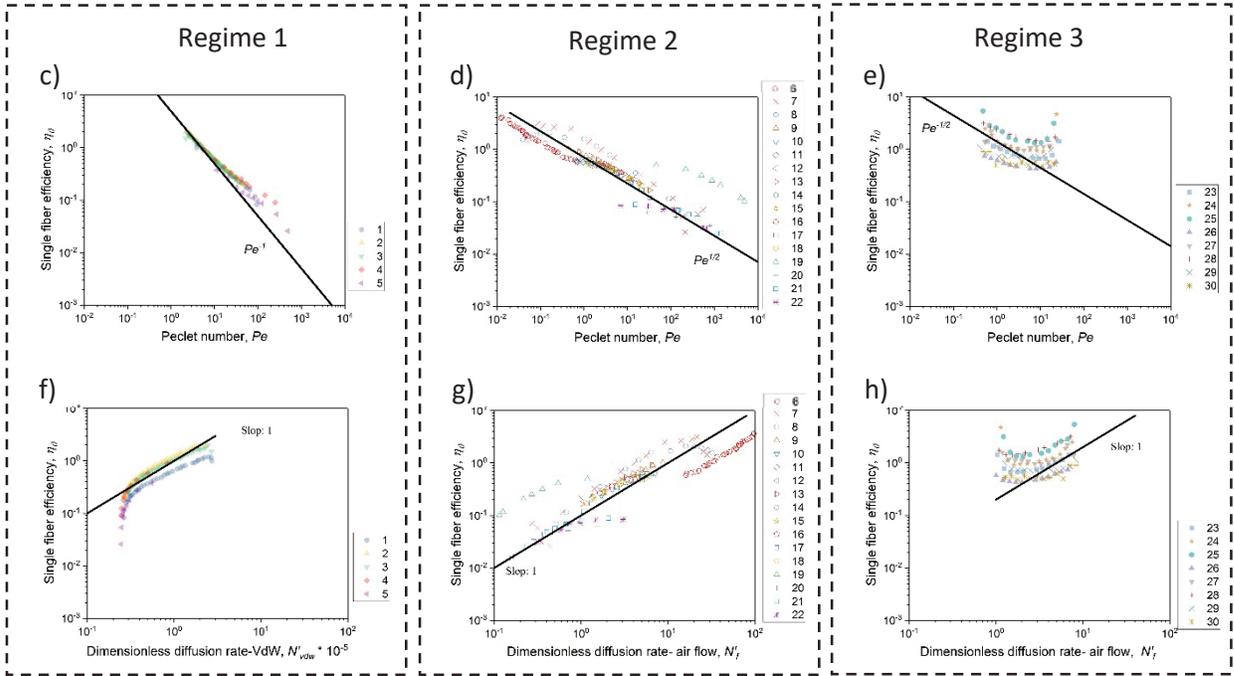

**Figure 2.** The single fiber efficiencies from experimental data at different Pe numbers and diffusion rates in three regimes

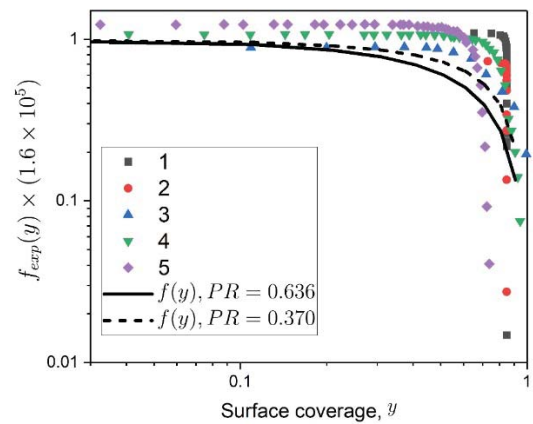

**Figure 3.** The surface coverage function of the model and experimental data

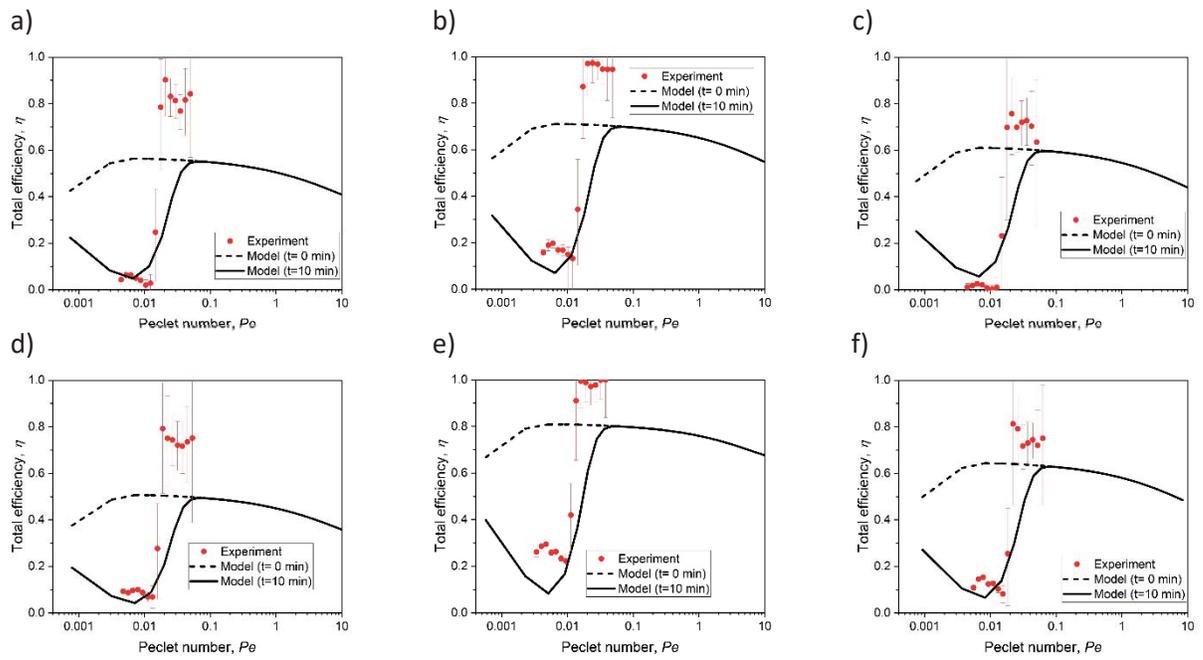

**Figure 4.** Modeled and experimental filtration efficiencies of WOx nanoparticles using PVA nanofibrous filters (a-f) 1-6 cases.

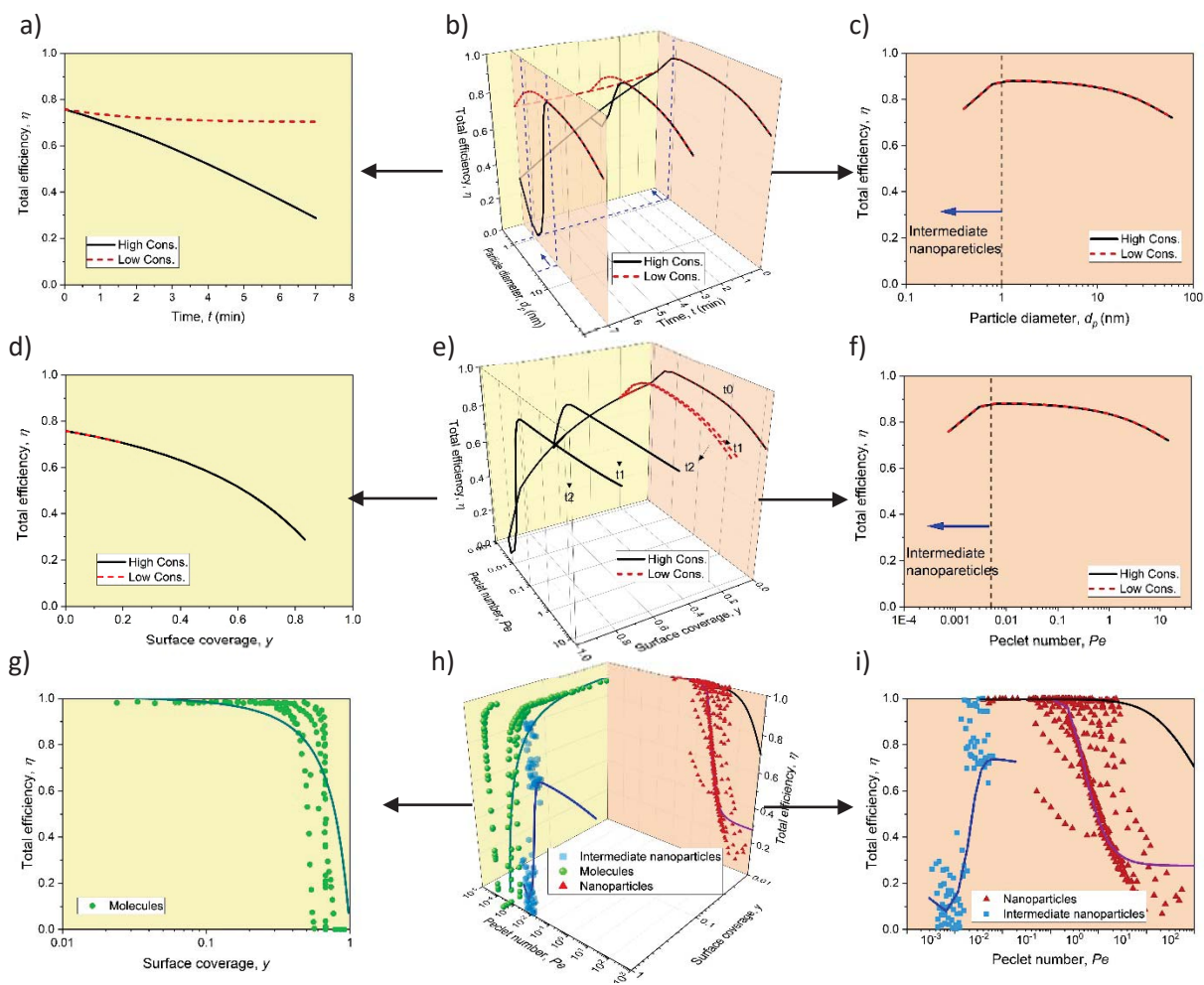

**Figure 5.** The unified model and its validation with experimental data, the total efficiency of the model (a-c) as a function of particle size and time, (d-f) as a function of Peclet number and surface coverage, and the total efficiency of experimental data (g-i) as a function of Peclet number and surface coverage

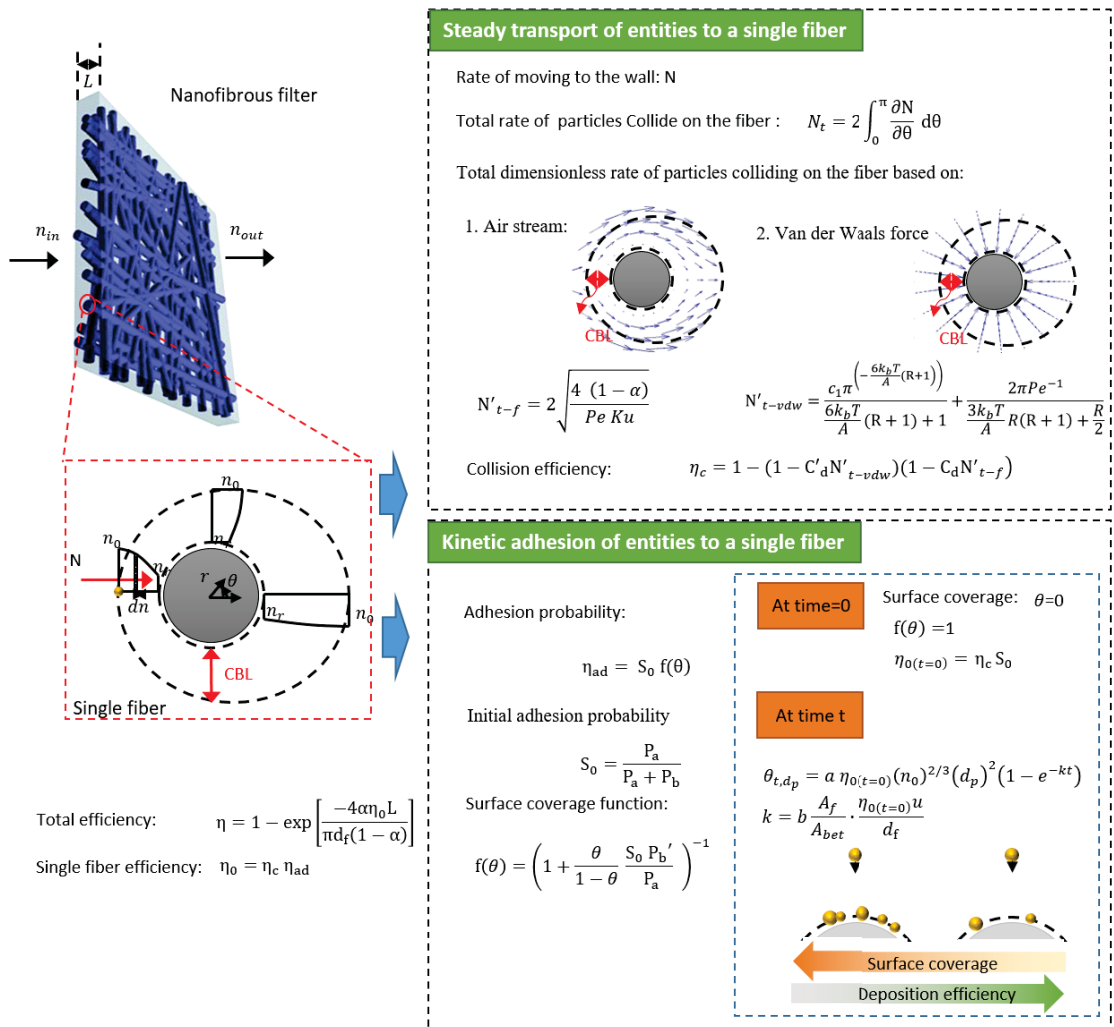

**Figure 6.** Model development bases on the steady transport of entities and kinetic adhesion

**This PDF file includes:**

    Supplementary text S1 to S9
    Figures S1 to S7
    Tables S1 to S3
    SI References

## List of Symbols

| Notation | Term | Unit |
|---|---|---|
| $A$ | Hamaker constant | J |
| $A_{bet}$ | Specific surface area | m²/g |
| $C_c$ | Slip correction factor | - |
| $d_f$ | Fiber diameter | m |
| $d_p$ | Particle diameter | m |
| $D$ | Diffusion coefficient of particles | m²/s |
| $E_{int}$ | Interaction energy | J |
| $e$ | Unit of charge | C |
| $F_{vdw}$ | Van der Waals force | N |
| $f_c$ | Charge fraction | - |
| $k$ | Overall effective mass transfer coefficient | 1/s |
| $k_b$ | Boltzmann's constant | J/K |
| $Kn_p$ | Particle Knudsen number | - |
| $Ku$ | Kuwabara hydrodynamic factor | - |
| $L$ | Thickness of filter | m |
| $L_{eff}$ | Effective length | m |
| $LOD$ | Limit of detection | #/m³ |
| m | Filter weight | g |
| $n$ | Particle number concentration | #/m³ |
| $n_0$ | Particle number concentration outside CBL | #/m³ |
| $n_r$ | Particle number concentration at fiber surface | #/m³ |
| $N$ | The rate of particles entering boundary layer in each θ | #/s |
| $N_t$ | Total rate of particles entering boundary layer | #/s |
| $N_f$ | The rate of particles entering boundary layer in each θ from air flow | #/s |
| $N_{vdw}$ | The rate of particles entering boundary layer in each θ from van der Waals foce | #/s |
| $Pe$ | Peclet number | - |
| $P_i$ | Probability of state i | - |
| $P_a$ | Probability of adhesion on an empty site | - |
| $P_b$ | Probability of bouncing from an empty site | - |

| $P_c$ | Probability of proceeding to the next site from an empty site | - |
| $P_a'$ | Probability of adhesion on an filled site | - |
| $P_b'$ | Probability of bouncing from an filled site | - |
| $P_c'$ | Probability of proceeding to the next site from an filled site | - |
| $P$ | Overall detection efficiency | |
| $q$ | Mass concentration at the surface at time t | - |
| $q_e$ | Maximum mass concentration at the surface | - |
| Q | Aerosol floe rate | m³/s |
| r | Polar coordinates component | m |
| $R_f$ | Fiber reduce | m |
| $R_p$ | Particle reduce | m |
| $S_0$ | Initial adhesion probability | - |
| $t$ | Time | s |
| $T$ | Temperature | K |
| $u$ | Aerosol face velocity | m/s |
| $u_r$ | Velocity component in polar coordinates | m/s |
| $u_\theta$ | Velocity component in polar coordinates | m/s |
| $v_{vdw}$ | Van der Waals velocity | m/s |
| $y$ | Surface coverage | - |
| $z_0$ | Equilibrium or minimum distance between bodies | m |
| $Z_p$ | Particle electrical mobility | m |

**Greek letters**

| α | Solidity of filter | - |
| $\beta$ | Mobility of particles | - |
| $\beta_r$ | Aerosol to sheath flow ratio | - |
| $\eta$ | Total filtration efficiency | - |
| $\eta_0$ | Single fiber filtration efficiency | - |
| $\eta_{ad}$ | Adhesion probability | - |
| $\eta_c$ | Collision efficiency | - |
| $\eta_{samp}$ | Penetration efficiencies through sampling line | - |
| $\eta_{char}$ | Penetration efficiencies through the charger | - |
| $\eta_{DMA}$ | Penetration efficiencies through DMA | - |
| $\eta_e$ | Penetration efficiencies through electrometer | - |
| $\mu_{air}$ | Kinetic viscosity of air | Kg/(m.s) |
| $\mu$ | Dimensionless deposition parameter | - |
| $\psi$ | Stream function | m²/s |
| $\psi_f$ | Stream function of air flow around nanofiber | m²/s |
| $\psi_{vdw}$ | Stream function of air flow around nanofiber | m²/s |
| $\rho$ | CBL thickness | m |
| $\rho_p$ | Particle density | Kg/ m³ |
| $\rho_f$ | Fiber density | Kg/ m³ |
| $\theta$ | Polar coordinates component | - |
| $\lambda$ | Mean free path | m |

## S1 Transformation of equation (2)

With a thin boundary layer around the fiber, the third term on the right-hand side of Equation (2) can be ignored. Therefore, it is reduced to:

$$u_r \frac{\partial n}{\partial r} + \frac{u_\theta}{r} \frac{\partial n}{\partial \theta} = \frac{D}{r} \frac{\partial}{\partial r} \left( r \frac{\partial n}{\partial r} \right) \tag{S-1}$$

where $D$ is the diffusion coefficient:

$$D = \frac{k_b T}{3 \pi d_p \mu_{air} C_c} \tag{S-2}$$

For nanoparticles the slip correction factor $C_c$ is based on Cunningham correlation in the gas flow (1):

$$C_c = 1 + Kn_p \left[ 1.257 + 0.4 \exp\left( -\frac{1.1}{Kn_p} \right) \right]^{1/3} \tag{S-3}$$

where $Kn_p$ is the particle Kundesn number ($Kn_p = 2\lambda/d_p$). Apply the chain rule and transform the coordinate system from $(r, \theta)$ to $(\psi, \theta)$:

$$\frac{\partial n}{\partial \theta} = \frac{\partial n}{\partial \psi} \frac{\partial \psi}{\partial \theta} + \frac{\partial n}{\partial \theta} \frac{\partial \theta}{\partial \theta} = r u_r \frac{\partial n}{\partial \psi} + \frac{\partial n}{\partial \theta} \tag{S-4}$$

$$\frac{\partial n}{\partial r} = \frac{\partial n}{\partial \psi} \frac{\partial \psi}{\partial r} + \frac{\partial n}{\partial \theta} \frac{\partial \theta}{\partial r} = -u_\theta \frac{\partial n}{\partial \psi} \tag{S-5}$$

$$\frac{\partial}{\partial r} \left( r \frac{\partial n}{\partial r} \right) = \frac{\partial}{\partial \psi} \left( r \frac{\partial n}{\partial r} \right) \frac{\partial \psi}{\partial r} = -u_\theta \frac{\partial}{\partial \psi} \left( r \frac{\partial n}{\partial r} \right) \tag{S-6}$$

Substituting (S-4), (S-5), and (S-6) into Equation (S-1) gives:

$$u_r \left( -u_\theta \frac{\partial n}{\partial \psi} \right) + \frac{u_\theta}{r} \left( r u_r \frac{\partial n}{\partial \psi} + \frac{\partial n}{\partial \theta} \right) = -u_\theta \frac{D}{r} \frac{\partial}{\partial \psi} \left( r \frac{\partial n}{\partial r} \right) \tag{S-7}$$

and

$$\frac{\partial n}{\partial \theta} = -D \frac{\partial}{\partial \psi} \left( r \frac{\partial n}{\partial r} \right) \tag{S-8}$$

Integrating both sides of Equation (S-7) with respect to $\psi$ gives:

$$\int \frac{\partial n}{\partial \theta} d\psi = -\int D \frac{\partial}{\partial \psi} \left( r \frac{\partial n}{\partial r} \right) d\psi \tag{S-9}$$

According to Leibniz integral rules:

$$\frac{\partial}{\partial \theta} \int_0^{\psi_\rho(\theta)} n(\psi, \theta) \, d\psi = n_0 \frac{\partial \psi_\rho(\theta)}{\partial \theta} - n_r \frac{\partial \psi_{Rf}(\theta)}{\partial \theta} + \int_0^{\psi_\rho(\theta)} \frac{\partial n}{\partial \theta} d\psi \tag{S-10}$$

where $\psi_{Rf}(\theta) = 0$. LHS of Equation (S-9) is replaced with Equation (S-10):

$$\frac{\partial}{\partial \theta} \int_0^{\psi_\rho(\theta)} n(\psi, \theta) d\psi - n_0 \frac{\partial \psi_\rho(\theta)}{\partial \theta} = -\int D \frac{\partial}{\partial \psi}\left(r \frac{\partial n}{\partial r}\right) d\psi \qquad \text{(S- 11)}$$

RHS of Equation (S- 11) is replaced with:

$$-\int D \frac{\partial}{\partial \psi}\left(r \frac{\partial n}{\partial r}\right) d\psi = -D\left[\left(\rho(\theta) \frac{\partial n}{\partial r}\right)_{\psi_\rho(\theta)} - \left(R_f \frac{\partial n}{\partial r}\right)_{R_f}\right] \qquad \text{(S- 12)}$$

With the CBL $\left(\frac{\partial n}{\partial r}\right)_{\psi_\rho(\theta)} = 0$, Equation (S- 11) becomes

$$n_0 \frac{\partial \psi_\rho(\theta)}{\partial \theta} - \frac{\partial}{\partial \theta} \int_{n_r}^{n_0} \psi dn - n_0 \frac{\partial \psi_\rho(\theta)}{\partial \theta} = D\left(R_f \frac{\partial n}{\partial r}\right)_{R_f} \qquad \text{(S- 13)}$$

$$-\frac{\partial}{\partial \theta} \int_{n_r}^{n_0} \psi dn = DR_f \left(\frac{\partial n}{\partial r}\right)_{R_f}$$

Define dimensionless parameters $\rho' = \frac{\rho}{R_f}$, $n' = \frac{n-n_r}{n_0 - n_r}$, $r' = \frac{r}{R_f}$, and $\psi' = \frac{\psi}{u R_f}$, and the dimensionless form of Equation (S- 13) is given as:

$$-\frac{\partial}{\partial \theta} \int_0^1 \psi' dn' = \frac{D}{u R_f} \frac{dn'}{dr'}\bigg|_{r' = 1} \qquad \text{(S- 14)}$$

**S2 Derivation of collision rate based on airflow**

The contribution of airflow to the rate of particles moving to the surface is:

$$N'_f = \int_0^1 \psi'_f \, dn' \qquad \text{(S- 15)}$$

The stream function of airflow over the fiber, $\psi_f$, is estimated with the Kuwabara flow field by considering randomly-distributed fibers (2). The dimensionless stream function is given as:

$$\psi'_f = \frac{1-\alpha}{Ku \, r'} (r' - 1)^2 \sin(\theta) \qquad \text{(S- 16)}$$

where Ku is the Kuwabara hydrodynamic factor:

$$Ku = -0.5 \ln\alpha - 0.75 - 0.25\alpha^2 + \alpha \qquad \text{(S- 17)}$$

Substituting the dimensionless stream function of Kuwabara flow (Equation (S- 16)) into Equation (S- 15) gives:

$$N'_f = \int_0^1 \frac{1-\alpha}{Ku \, r'} (r' - 1)^2 \sin(\theta) \, dn' \qquad \text{(S- 18)}$$

$$= \frac{(1-\alpha)\sin(\theta)}{Ku} \int_0^1 \frac{(r'-1)^2}{r'} dn'$$

The concentration gradient near the fiber surface can be estimated using Fick's second law (i.e., $\frac{dn'}{dr'} = \frac{1}{r' Ln(\rho')}$). Thus Equation (S- 18) becomes:

$$N'_f = \frac{(1-\alpha)\sin(\theta)}{Ku\ \ln(\rho')} \int_0^1 \frac{(r'-1)^2}{r'} \frac{1}{r' Ln(\rho')} dr' \qquad \text{(S- 19)}$$

$$= \frac{(1-\alpha)\sin(\theta)}{Ku\ \ln(\rho')} \left[\rho' - \frac{1}{\rho'} - 2Ln(\rho')\right]$$

The logarithm term can be approximated by the Taylor series expansion while a variable, x, is close to unity:

$$Ln\ x = (x-1) + \frac{(x-1)^2}{2} + \frac{(x-1)^3}{3} + \cdots$$

Or

$$Ln\ x = \left(1-\frac{1}{x}\right) + \frac{1}{2}\left(1-\frac{1}{x}\right)^2 + \frac{1}{3}\left(1-\frac{1}{x}\right)^3 + \cdots$$

Therefore, $Ln\ (\rho')$ is approximated as $(\rho'-1)$ or $\left(1-\frac{1}{\rho'}\right)$. Then:

$$N'_f = -\frac{(1-\alpha)\sin(\theta)}{Ku} \cdot \frac{(\rho'-1)}{\rho'} \qquad \text{(S- 20)}$$

and

$$\frac{(\rho'-1)}{\rho'} = -\frac{N'_f\ Ku}{(1-\alpha)\sin(\theta)} \qquad \text{(S- 21)}$$

Substituting the concentration gradient into Equation (S- 14) gives:

$$-\frac{\partial N'}{\partial \theta} = \frac{D}{u\ R_f} \frac{1}{Ln(\rho')} \qquad \text{(S- 22)}$$

The term of $Ln\ (\rho')$ in Equation (S- 21) is replaced from Equation (S- 22), thus:

$$\frac{N'_f\ Ku}{(1-\alpha)\sin(\theta)} \frac{\partial N'_f}{\partial \theta} = \frac{D}{u\ R_f} \qquad \text{(S- 23)}$$

Rearrangement of Equation (S- 23) gives:

$$N'_f\ dN'_f = \frac{D\ (1-\alpha)\sin(\theta)}{u\ R_f Ku} d\theta \qquad \text{(S- 24)}$$

Integration of both sides gives:

$$\frac{N'^2_f}{2} = -\frac{D\ (1-\alpha)}{Ku.u\ R_f} \cos(\theta) + c_0 \qquad \text{(S- 25)}$$

Assuming a negligible rate at $\theta = 0$, we can determine the constant $c_0$:

$$c_0 = \frac{D(1-\alpha)}{u\,R_f Ku} \tag{S-26}$$

The dimensionless rate of particles moving to the fiber's surface is calculated as:

$$N'_f = \sqrt{\frac{4D(1-\alpha)}{u\,R_f\,Ku}[1-\cos(\theta)]} \tag{S-27}$$

The gradient of this quantity shows the rate of particles that collide on the fiber surface. Total dimensionless rate of particles colliding on the surface is determined by integration the gradient of rate for particle movement to the surface ($\frac{\partial N'}{\partial \theta}$) integral over the fiber's surface gives the total dimensionless gradient of the rate of moving to the unit length of the fiber.

$$N'_{t-f} = 2\int_0^\pi \frac{\partial N'_f}{\partial \theta}\,d\theta \tag{S-28}$$

$$= 2\sqrt{\frac{4(1-\alpha)}{Pe\,Ku}}$$

**S3 Derivation of collision rate based on van der Waals force**

Contribution of van der Waals force to the rate of particles moving to the surface is:

$$N'_{vdw} = \int_0^1 \psi'_{vdw}\,dn' \tag{S-29}$$

The stream function due to the van der Waals is determined by velocity components $u_r = -v_{vdw}$ and $u_\theta = 0$:

$$-v_{vdw} = \frac{1}{r}\frac{\partial \psi_{vdw}}{\partial \theta} \tag{S-30}$$

$$0 = -\frac{\partial \psi_{vdw}}{\partial r} \tag{S-31}$$

The stream function is determined using Equations (S- 30) and (S- 31) as:

$$\psi_{vdw} = \int -r\,v_{vdw}\,d\theta + f(r) \tag{S-32}$$

$$= -r v_{vdw}\,\theta + c_0$$

At $r = R_f$, $\psi_{vdw} = 0$, then Equation (S- 32) becomes:

$$\psi_{vdw} = -r v_{vdw}\,\theta + R_f\,v_{vdw}|r = R_f\,\theta \tag{S-33}$$

van der Waals velocity, $v_{vdw}$, is the product of particle mobility and van der Waals force acting on the particle.

$$v_{vdw} = \beta F_{vdw} \tag{S-34}$$

where $\beta$ is the mobility of particles ($\beta = \frac{C_c}{3\pi\mu_{air}d_p}$). $F_{vdw}$ between two bodies with radiuses of $R_f$ and $R_p$ is given by (3):

$$F_{vdw} = \frac{A}{6[r-(R_p+R_f)]^2}\left(\frac{R_p R_f}{R_p+R_f}\right) \tag{S-35}$$

Substituting $v_{vdw}$ from Equations (S-34) and (S-35) and dimensionless quantities into (S-33), the dimensionless form of stream function due to van der Waals is obtained as:

$$\psi'_{vdw} = \frac{\beta A \theta}{6uR_f}\left(\frac{R_p R_f}{R_p+R_f}\right)\left[\frac{-r'}{\left(r'-\left(\frac{R_p}{R_f}+1\right)\right)^2}+\left(\frac{R_f}{R_p}\right)^2\right] \tag{S-36}$$

and

$$\psi'_{vdw} = A_1 \theta \left[\frac{-r'}{\left(r'-\left(\frac{R_p}{R_f}+1\right)\right)^2}+\left(\frac{R_f}{R_p}\right)^2\right] \tag{S-37}$$

where $A_1$ is $\frac{\beta A}{6uR_f}\left(\frac{R_p R_f}{R_p+R_f}\right)$. Substituting stream function of Equation (S-37) into Equation (S-29) gives:

$$N'_{vdw} = \int_0^1 A_1 \theta \left[\frac{-r'}{\left(r'-\left(\frac{R_p}{R_f}+1\right)\right)^2}+\left(\frac{R_f}{R_p}\right)^2\right] dn' \tag{S-38}$$

Equation (S-38) is transformed into:

$$\tag{S-39}$$
$$N'_{vdw} = \int_0^1 A_1 \theta \left[\frac{-r'}{\left(r'-\left(\frac{R_p}{R_f}+1\right)\right)^2}+\left(\frac{R_f}{R_p}\right)^2\right]\frac{1}{r'Ln(\rho')}dr'$$
$$= \frac{A_1 \theta}{Ln(\rho')}\left[\frac{R_f}{R_p}+\left(\frac{R_f}{R_p}\right)^2 Ln(\rho')\right]$$

$$= \frac{A_1 \theta}{Ln(\rho')} \left(\frac{R_f}{R_p}\right) + A_1 \theta \left(\frac{R_f}{R_p}\right)^2$$

The term $\frac{1}{Ln(\rho')}$ is obtained by rearranging Equation (S- 39):

$$\frac{1}{Ln(\rho')} = \frac{N'_{vdw}}{A_1 \theta \left(\frac{R_f}{R_p}\right)} - \left(\frac{R_f}{R_p}\right) \quad \text{(S- 40)}$$

Equation (S- 14) then becomes:

$$-\frac{\partial}{\partial \theta} N'_{vdw} = \frac{D}{u\, R_f} \left[\frac{N'_{vdw}}{A_1 \theta \left(\frac{R_f}{R_p}\right)} - \left(\frac{R_f}{R_p}\right)\right] \quad \text{(S- 41)}$$

Equation (S- 41) is an ordinary differential Equation (ODE) in the form of :

$$\frac{\partial N'_{vdw}}{\partial \theta} = -a\, \frac{N'_{vdw}}{\theta} + b \quad \text{(S- 42)}$$

where a and b are:

$$a = \frac{D}{u\, R_f A_1 \left(\frac{R_f}{R_p}\right)} \quad \text{(S- 43)}$$

$$b = \frac{D}{u\, R_p} \quad \text{(S- 44)}$$

ODE Equation (S- 42) is solved by changing variable $\frac{N'_{vdw}}{\theta} = u'(\theta)$. Then the first derivative is:

$$\frac{\partial N'_{vdw}}{\partial \theta} = \frac{\partial u'}{\partial \theta} \theta + u' \quad \text{(S- 45)}$$

With the new variable, Equation (S- 42) becomes:

$$\frac{\partial N'_{vdw}}{\partial \theta} = -a\, u' + b \quad \text{(S- 46)}$$

Combining both Equations (S- 45) and (S- 46) and rearranging them gives:

$$-\frac{d\theta}{\theta} = \frac{du'}{(a+1)u' - b} \quad \text{(S- 47)}$$

Equation (S- 47) is solved as:

$$-Ln(\theta) + c_0 = \frac{Ln\left[(a+1)u' - b\right]}{a+1} \tag{S-48}$$

Thus, by replacing $N'_{vdw}/\theta = u'(\theta)$ into Equation (S-48), the dimensionless rate of particles moving to the fiber's surface is calculated as:

$$N'_{vdw} = \frac{\theta\left(c_1\theta^{-(a+1)} + b\right)}{a+1} \tag{S-49}$$

The total dimensionless rate of particles colliding on the surface of a single fiber per unit length is:

$$N'_{t-vdw} = 2\int_0^\pi \frac{\partial N'_{vdw}}{\partial \theta}\, d\theta = \frac{2\pi\left(c_1\pi^{-(a+1)} + b\right)}{a+1} \tag{S-50}$$

Calculated dimensionless rate of collision due to the van der Waals from our model is rearranged as a function of $Pe$ and R:

$$N'_{t-vdw} = \frac{c_1\pi^{\left(-\frac{6k_bT}{A}(R+1)\right)}}{\frac{6k_bT}{A}(R+1) + 1} + \frac{2\pi Pe^{-1}}{\frac{3k_bT}{A}R(R+1) + \frac{R}{2}} \tag{S-51}$$

**S4 Derivation of adhesion probability**

As the size of particles decreases to the intermediate size range, the non-perfect assumption is reasonable due to the similarities between intermediate nanoparticle and molecules. With the non-perfect assumption, only a fraction of those particles that collide on the surface can adhere to the surface. Thus, the product of collision efficiency $\eta_c$ from Equation (7) and adhesion probability $\eta_{ad}$ demonstrates the efficiency of a single fiber, $\eta_0$:

$$\eta_0 = \eta_{ad}\eta_c \tag{S-52}$$

Although there is no clear definition for a contact of two surfaces at the molecular level, the definition seems reasonable based on the earlier studies (4, 5): intermediate particles are in contact with the surface when the atoms are closer than $r_e$ (the separation at the minimum potential energy). At this separation distance, the intermediate nanoparticles will be at equilibrium. No van der Waals force exerted on the particle ($F_{int} = 0$) (4, 5), although all particles on the surface or air stream are under thermal motion and might depart from the surface when their energy overcomes this potential energy. Among various adhesion models considering particles and a surface at the contact (6-10), the Hamaker model is the closest to this definition. Therefore, the Hamaker model is used in this work to model the f particle-surface and particle-particle interactions. The contact area around the fiber is shown in Figure 1a. Discreetness of intermediate nanoparticles, non-additivity, and retardation are ignored (11, 12).

Assuming intermediate particles behave like gas molecules, the adhesion probability of the intermediate size range is determined. Figure S1 illustrates the similarity between molecules and intermediate nanoparticles when the adhesive van der Waals force from the surface exerted on them. The intermediate particles are stabilized with a tiny separation distance, $z_0$, where the particle is balanced by adhesive and repulsive interaction forces. Thus, intermediate particles are in contact with the surface, while it is the separation shorter than $z_0$. However, particles adhere to the surface, although they are in physical contact with it. One of the main differences between molecules and particles in the interfacial interaction is the repulsive force. The pair potential stabilizes molecules on the surface; however, for particles, the repulsive force from the elastic/plastic deformations leads to their stabilizing on the surface (12). Intermediate particles can oscillate in a separation distance shorter than $z_0$ because of the energy exchange. The intermediate particle would dissociate from the surface when it gains enough energy.

The intermediate nanoparticles gain energy from frequent thermal or Brownian motion of the surrounding gas molecules, whether the particles are in the air or on the surface (13). While they are in the air, this thermal motion leads to particle diffusion (14). The adhesion probability of nanoparticles is:

$$\eta_{ad} = S_0 \, f(y) \tag{S-53}$$

where $S_0$ is the initial adhesion probability, $f(y)$ is surface coverage function, and $y$ is the dimensionless surface coverage. The initial adhesion probability and surface coverage function are determined by considering all probable kinetic processes: adhesion, detaching, and proceeding to the next site, which are denoted as $P_a$, $P_b$, and $P_c$, respectively (Figure S2). Additionally, $P_a'$, $P_b'$ and $P_c'$ denote the probabilities of adhesion, detaching, and proceeding to the next site when the particle strikes on a site that is already occupied (15). Thus, the probability of adhesion, bouncing and moving to the next site on the first site are:

$$P_{a1} = (1-y)P_a \tag{S-54}$$

$$P_{b1} = (1-y)P_b + y\,P_b' \tag{S-55}$$

$$P_{c1} = 1 - P_{a1} - P_{b1} \tag{S-56}$$

With the assumption that the probabilities for the first site are independent on others, the probabilities on the second site are:

$$P_{a2} = P_{c1}\big((1-y)P_a\big) \tag{S-57}$$

$$P_{b2} = P_{c1}\big((1-y)P_b + y\,P_b'\big) \tag{S-58}$$

$$P_{c2} = P_{c1}(1 - P_{a1} - P_{b1}) \tag{S-59}$$

The sum of probabilities of adhesion over all available sites defines the adhesion probability as:

$$\eta_{ad} = P_{a1} + P_{a2} + P_{a3} + \cdots \tag{S-60}$$

The initial adhesion coefficient, $S_0$, is defined as the ratio of the number of adsorbed particles to that of impinging particles at zero coverage.

$$S_0 = \frac{P_a}{P_a + P_b} \tag{S-61}$$

Assuming monolayer adsorption ($P_a' = 0$), the surface coverage function is:

$$f(y) = \left(1 + \frac{y}{1-y}\frac{S_0\,P_b'}{P_a}\right)^{-1} \tag{S-62}$$

In the case of multilayer deposition, the description of the Kisliuk model can be revised. For multilayer assumption, $P_a'$ is considered non-zero because intermediate nanoparticles can adhere to the occupied site. Then, the probability of adhesion on the first site is defined as:

$$P_{a1} = (1-y)P_a + y\,P_a' \tag{S-63}$$

With the assumption that the probabilities for the first site are independent on other sites, the probability of adhesion on the second site is:

$$P_{a2} = P_{c1}[(1-y)P_a + y\,P_a'] \tag{S-64}$$

The probability of detaching from and adhesion onto the second site are described using Equations (S- 61) and (S- 62). The sum of probabilities of adhesion over all available sites becomes a new function for the adhesion probability.

$$\eta_{ad} = P_{a1} + P_{a2} + P_{a3} + \cdots = \left((1-y)P_a + y\,P_a'\right)\left(1 + P_{c1} + P_{c1}^2 + \cdots\right) \tag{S-65}$$

$$= \frac{(1-y)P_a + y\,P_a'}{1 - P_{c1}}$$

$$= \frac{P_a + y\,(P_a' - P_a)}{P_a + P_b + \theta\,(P_a' - P_a + P_b' - P_b)}$$

Rearranging Equation (S- 65) gives:

$$\eta_{ad} = \frac{P_a}{P_a + P_b} \frac{1 + ny}{1 + my} \tag{S-66}$$

where $n = \frac{P_a'}{P_a} - 1$ and $m = \frac{P_a' + P_b'}{P_a + P_b} - 1$. For zero surface coverage (y=0), the adhesion efficiency is reduced to the initial adhesion probability. The probabilities of different states are estimated based on the partition function.

The adhesion to the surface results from the interfacial interaction of van der Waals. Since physisorbed intermediate nanoparticles are at thermal equilibrium with respect to the surface (15), the probability of adhesion can be determined with the partition function (16). Each particle represents a system, and particles in the same state have equal energy and probability. The ensemble moves towards the state that has the maximum entropy. The probability of state i out of $\Omega$ states is then obtained by the Lagrange multipliers (16):

$$P_i = \frac{\exp(-E_i/k_b T)}{\sum_1^\Omega \exp(-E_i/k_b T)} \tag{S-67}$$

where $E_i$ is the energy of a system in state i. By determining the probabilities of six possible kinetic states, the initial adhesion probability and the surface coverage function are determined from all the states defined earlier.

Interaction potentials between bodies due to van der Waals force equal to the energy of adsorption, or the difference between energies of the entities in the adsorbed and the gas state (17). The difference between $E_a$ and $E_b$ is the particle–surface interfacial interaction, whereas that between $E_a'$ and $E_b'$ is particle-particle interfacial interaction due to the van der Waals. Since the energy barrier for detachment of intermediate particle in multilayer deposition ($E_{pp}$) is lower than that in the monolayer ($E_{sp}$), the probability of adhesion in the occupied site is lower than a clean site. With the assumption of the attractive interaction potential between two atoms or small molecules with a separation distance of $z_0$ ($E_{int}(r) = -C/r^n$), the interaction potential between two entities is the sum of the interaction energies among all molecules in the two bodies of concern. Hamaker (10) derived a widely-accepted equation for van der Waals interaction between two spheres for all separations:

$$E_{int} = -\frac{A}{6}\left\{\frac{2R_1 R_2}{(2R_1 + 2R_2 + z_0)z_0} + \frac{2R_1 R_2}{(2R_1 + z_0)(2R_2 + z_0)} + \ln\frac{(2R_1 + 2R_2 + z_0)z_0}{(2R_1 + z_0)(2R_2 + z_0)}\right\} \tag{S-68}$$

where $R_1$ and $R_2$ are radii of the two bodies, $z_0$ is the separation distance, and A is the effective Hamaker constant for interacting bodies 1 and 2 ($A = \sqrt{A_1 A_2}$).

**S5 Derivation of surface coverage**

Concentration and time may have an interchangeable role on the surface coverage. Over time and in high concentrations the fiber surface is covered with more particles, which reduces the adhesion probability and deposition efficiency. Experiments also show the deposition of intermediate nanoparticles is concentration dependant for a short time of operation (18). For adsorption, breakthrough curves that show the ratio of downstream to upstream concentration versus time are also concentration-dependent.

Adsorption kinetics can be modeled as a diffusion control process based on the linear film diffusion model (LDM) (19):

$$\frac{dq}{dt} = k\,(q_e - q) \tag{S-69}$$

where $q$ and $q_e$ are mass concentration and potential maximum mass concentration at the surface, respectively; $k$ is the overall effective mass transfer coefficient. There is no rigid calculation available for this coefficient. The LDF mass transfer coefficient has been calculated from experimental breakthrough fronts (20). The correlation between the mass transfer coefficient to the effective diffusion coefficient depends on the geometry of the adsorbent (20, 21), but it is constant with time. The coefficient $k$ for all entities is also scaled with the ratio of dimensionless bet specific surface area of the filter:

$$k = b\,\frac{A_f}{A_{bet}} \cdot \frac{\eta_{0(t=0)} u}{d_f} \tag{S-70}$$

Integrating Equation (S- 69) for boundary conditions t=0 to t and q=0 to $q$, repectivley, gives:

$$q = q_e(1 - e^{-kt}) \tag{S-71}$$

The surface coverage, y, is obtained by the correlation between mass concentration and surface coverage. Starting from t=0, the number of particles with the size $d_p$ deposited on the single fiber is $(n_0)^{2/3}\eta_{0(t=0)}$ (#/cm²). Thus, the mass concentration at the surface is determined at t=0 by multiplying number concentration and mass of particles with the size $d_p$:

$$q_{(t=0,d_p)} = \frac{\pi(d_p)^3}{6}\rho_p\,(n_0)^{2/3}\eta_{0(t=0)} \tag{S-72}$$

The total mass concentration on the fiber at time t is:

$$q_{(d_p)} = \int_0^t \frac{\pi(d_p)^3}{6}\rho_p\,(n_0)^{\frac{2}{3}}\eta_{0(t)}\,dt$$
$$= \frac{\pi(d_p)^3}{6}\rho_p\,(n_0)^{\frac{2}{3}} \int_0^t \eta_{0(t)}\,dt \tag{S-73}$$

As the deposition during the time is reduced due to the lower efficiency, $\eta_{0(t)}$, the maximum deposition takes place at time t=0. Therefore, maximum mass concentration is assumed to be proportional to the deposition at t=0.

$$q_{e\,(d_p)} = b\,\frac{\pi(d_p)^3}{6}\rho_p\,(n_0)^{2/3}\eta_{0(t=0)} \tag{S-74}$$

where b is constant. For monolayer deposition, all particles with the size $d_p$ are deposited on the surface of the fiber. Similarly, the surface coverage is estimated as:

$$y_{d_p} = \int_0^t \frac{\pi(d_p)^2}{4} (n_0)^{\frac{2}{3}} \eta_{0(t)} \, dt$$
$$= \frac{\pi(d_p)^2}{4} (n_0)^{\frac{2}{3}} \int_0^t \eta_{0(t)} \, dt \qquad \text{(S- 75)}$$

Writing Equation (S- 71) for particles with the size $d_p$ and substituting Equations (S- 73), (S- 74) and (S- 75) into Equation (S- 71), we have:

$$y_{d_p} = b \, n_0^{2/3} \eta_{0(t=0)} \pi (d_p)^2 (1 - e^{-kt}) \qquad \text{(S- 76)}$$

Regarding the size dependency of $\eta_{0(t=0)}$ as well as concentration distribution in the polydisperse case, the total surfaces coverage, $y$, is obtained by integrating Equation (S- 76) over the size from 0 to $+\infty$:

$$y = \int_0^{+\infty} y_{t,d_p} \, d(d_p) = b \left( \int_0^{+\infty} n_0^{2/3} \eta_{0(t=0)} \pi (d_p)^2 \, dd_p \right) (1 - e^{-kt}) \qquad \text{(S- 77)}$$

**S6 The rate of diffusion due to the van der Waals**

Figure S3 shows the dimensionless rate of collision based on van der Waals as a function of $Pe$ and $R$. The lines represent the rates for three fiber diameters. For all fiber diameters, the rate relatively constant for high $R$ and $Pe$ values (larger particles). The green iso-surface shows the constant value of this rate. With decreasing $Pe$ number and increasing $R$ (intermediate particles), the rate begins to increase sharply after a specific point (shown with asterisks). All asterisks on the green solid line corresponded to the same particle size. This critical size is unique and depends on the operating parameters and properties. There is a critical particle size beyond which the dimensionless rate of collision based on van der Waals becomes negligible. The dimensionless rate of collision based on van der Waals is negligible for the data that deviates from the linear behavior in regime 1. These data are between regimes 1 and 2.

**S7 Equivalent properties of nanofibrous media**

The total filtration efficiency can be calculated using the classical filtration model proposed by Spumy and Pich (22):

$$\eta = 1 - \exp\left[\frac{-4\alpha \eta_0 L}{\pi d_f (1 - \alpha)}\right] \qquad \text{(S- 78)}$$

where $\eta_0$ is the single fiber efficiency, $\alpha$ is the filter solidity, $L$ is the thickness of the filter, and $d_f$ is the mean diameter of the fibers.

Equivalent filter properties can be used while applying Equation (S- 78) to the adsorption on nanofibers. The solidity is calculated from its definition:

$$\alpha = \frac{d_s \pi d_f^2}{4 A_f L} \qquad \text{(S- 79)}$$

where $d_s$ is the length of the uniform single fiber. On the other side, specific surface area, $A_{bet}$, is generally used in adsorption studies, and it is defined as the total surface area of the material per unit of mass. $A_{bet}$ for the fibers can be calculated as:

$$A_{bet} = \frac{d_s \pi d_f}{m} \qquad \text{(S- 80)}$$

Substituting $d_s$ from Equation (S- 79) into Equation (S- 80) gives:

$$L = \frac{d_f A_{bet} m}{4 A_f \alpha} \tag{S- 81}$$

Then we have

$$\frac{4\alpha L}{\pi d_f (1-\alpha)} = \frac{A_{bet} m}{\pi A_f (1-\alpha)} \tag{S- 82}$$

**S8 Evaluation of experimental data for the filtration of intermediate nanoparticles**

The experimental data from reference (18) was used to validate the modelThe systematic error related to the experimental data from literature is determined from limit of detection (LOD) of measuring device for sub-4nm particles. The concentration of particles from (18) is measured by SMPS+E. In comparison with CPCs, FCEs have higher counting efficiency due to the high diffusional loss for very small nanoparticles. Furthermore, FCEs are not sensitive to operating conditions, particle composition, and charge state (23-26). However, FCEs have a high background noise level. The noise level is described by the root mean square (RMS) of the signal. The limit of detection of FCE is different from CPC, it can be estimated based on the root mean square (RMS) value of the noise level based on the average over the measurement time:

$$LOD = \frac{3(\ln 10)\ RMS_1}{\pi_{FCE}} \tag{S- 83}$$

Coefficient 3 is needed for possible background count interference of one or two raw counts. The RMS based on the average measurement time is determined by the RMS over 1 s.

$$RMS_t = \frac{RMS_1}{\sqrt{t/1}} \tag{S- 84}$$

The parameter $\pi_{FCE}$ in Equation (S- 83) is defined as a product of the overall detection efficiency, the root of measurement time, the aerosol to sheath flow ratio, and the effective aerosol flow rate (27):

$$\pi_{FCE} = eP\sqrt{t/1s}\ Q\ \beta_r \left(-\frac{d \log d_p}{d \log Z_p}\right) \tag{S- 85}$$

where $e$ is a unit of charge, $Z_p$ is the particle electrical mobility, $\beta_r$ is aerosol to sheath flow ratio and $Q$ is the effective aerosol flow rate. The overall detection efficiency is usually determined by the product of efficiencies through the instrument:

$$P = \eta_{samp} \eta_{char} \eta_{DMA} f_c \eta_e \tag{S- 86}$$

where $f_c$ is the charge fraction of the particle, which is determined using the Fuchs theory. Based on this charge distribution, particles smaller than 13 nm accept only a single charge (28). In Equation (S-86), $\eta_{samp}$, $\eta_{char}$, $\eta_{DMA}$, and $\eta_e$ are the penetration efficiencies through sampling line, the charger, the DMA, and the electrometer. The overall penetration efficiency is:

$$\eta_{pen} = \eta_{samp} \eta_{char} \eta_{DMA} \tag{S- 87}$$

Based on the effective length approach, the particle loss through different parts of the instrument is estimated. The overall penetration efficiency is given by (29):

$$\eta_{pen} = 0.819 \exp(-3.66\mu) \\ + 0.0975 \exp(-22.3\mu) \\ + 0.0325 \exp(-57.0\mu) + 0.0154 \exp(-107.6\mu) \tag{S- 88}$$

$$\eta_{pen} = 1 - 2.56\,\mu^{\frac{2}{3}} + 1.2\,\mu + 0.1767\mu^{\frac{4}{3}} \quad \begin{array}{l} for\ \mu > 0.02 \\ for\ \mu \leq 0.02 \end{array}$$

where $D$ is the diffusion coefficient of particles, Q is the flow rate through the cylindrical tube, $L_{eff}$ is the effective length, and $\mu$ is the dimensionless deposition parameter:

$$\mu = \frac{\pi\,D\,L_{eff}}{Q} \tag{S-89}$$

**S9 Discussion on the adhesion probability of intermediate nanoparticles**

Figure S4 compares the model with an experimental study (30) for two cases (Figures S4a and S4b). The properties of the two cases are listed in Table S1. Measured total efficiencies are shown in scattered markers, where NaCl nanoparticles are deposited on a glass fibrous filters. Figure S4 shows that the dominant mechanism of transport is convective diffusion; thus, $C_d N'_f$ is a dominant term in Equation 7. The solid lines show the model results considering different correction factors. Solid red lines are for $C_d = 1/(1 + N'_f)$ from (31), and solid blue lines, for $C_d = 0.2$ using our model herein.

For the particles larger than 3 nm, the model considering $C_d = 0.2$ (blue line) is consistent with the experimental data in Figure s S4a and S4b; however, there is no efficiency drop. By considering $C_d = 0.2$, the collision efficiency for sub-3 nm is higher than unity, but the total efficiency is considered as 100 %. The model shows that the single-fiber efficiency increases with decreasing particle size due to the diffusion mechanism; however, the single-fiber efficiency begins to drop when the size is 3 nm or smaller. Since the initial adhesion probability is lower than unity for sub-3 nanoparticle, the adhesion of NaCl particles on the glass fibrous fibers is no longer 100%. Depending on the properties of filter, the total filtration efficiency might decrease for sub-3 nm particles. The vertical dashed line at 3 nm shows the onset of efficiency drop due to the lower adhesion probability.

The model with $C_d = 1/(1 + N'_f)$ shows that the efficiency drops for sub-2 nm particles, which is consistent with the experiment. For sub3 nm particles, the model result and the experimental data are following the same trend, although the experimental data has a 10 % vertical shift in Figure S4a and 6% in Figure S4b. It can be concluded that the correction factor of $C_d = 1/(1 + N'_f)$ should be considered for intermediate particles, otherwise, the correction factor is 0.2.

Figure S5 and Figure S6 show eight experimental data (32) and two more cases (33) compared to the model. The properties used are listed in Table S1. The dominant mechanism of transport for theses cases is convective diffusion, and $C_d N'_f$ is the dominant term in Equation 7. The results are depicted in solid red line for $C_d = 1/(1 + N'_f)$ (31), and the solid blue line, for $C_d = 0.2$.

The initial adhesion probability is lower than unity for sub-1.8 nm particles. Although the single fiber efficiency is expected to drop for this range, the model shows no efficiency drop with the filter properties used. For the particles larger than 1.8 nm, the model using $C_d = 0.2$ (blue line) agrees with the experimental data (see Figure S5 and Figure S6). For sub-1.8 nm particles, the model results considering $C_d = 1/(1 + N'_f)$ shows no efficiency drops, which is consistent with the experiment.

Figure S7 compares the model with the total efficiencies for 3 experimental data (34). Figure S7a and S7b are for the results with Tungsten oxide nanoparticles are deposited on stainless steel wires, and Figure S7c is for nickel particles. The solid red line is for $C'_d = 1/(1 + N'_{vdw})$ (31), and the solid blue line is for $C'_d = 10^{-5}$.

The experimental data show a slight efficiency drop starting from 1.3 nm. The model with $C_d = 0.2$ (blue line) agrees with the experimental data when the particles are larger than 1.8 nm. For sub-1.8 nm particles, the model results considering $C_d = 1/(1 + N'_f)$ shows no efficiency drops, which is consistent with the experiment. According to the model, the initial adhesion probability is lower than unity for sub-1.3 nanoparticle interacting with stainless steel and sub-1.4 nanoparticle interacting with nickel, however with respect to the filter's properties, the model results (blue line) shows the drop starting from 1.3 nm nanoparticle to the smaller sizes in all cases.

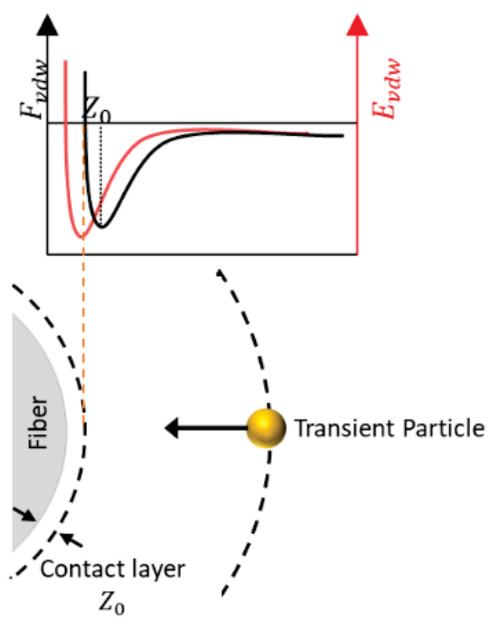

**Fig. S1.** The adhesion of intermediate nanoparticles to the fiber's surface.

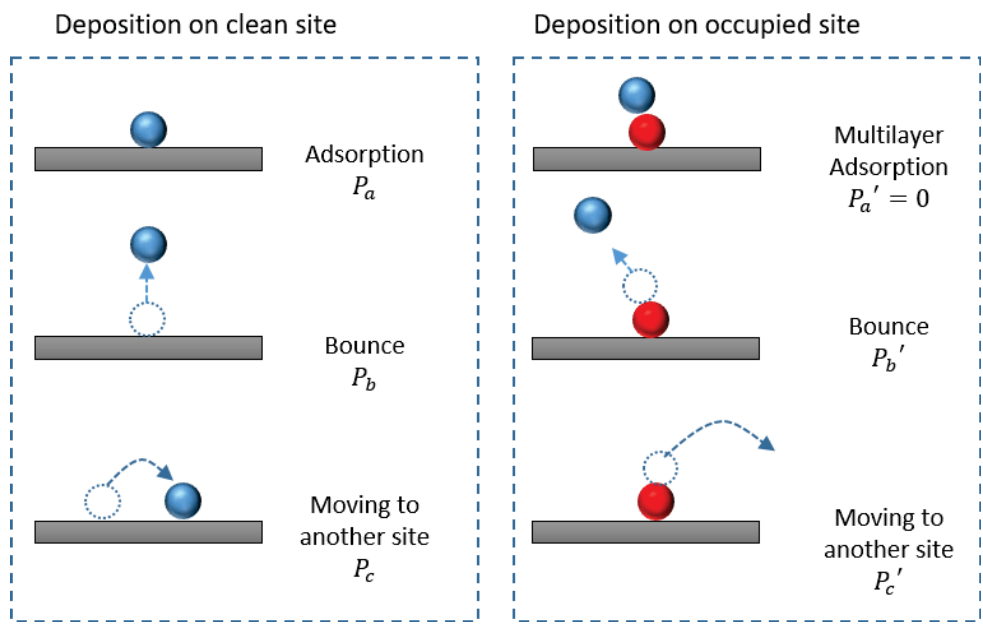

**Fig. S2.** The probabilities of adhesion, bounce, and moving to another site for an empty and occupied site (Blue particles are the colliding particles, and red, deposited particles.)

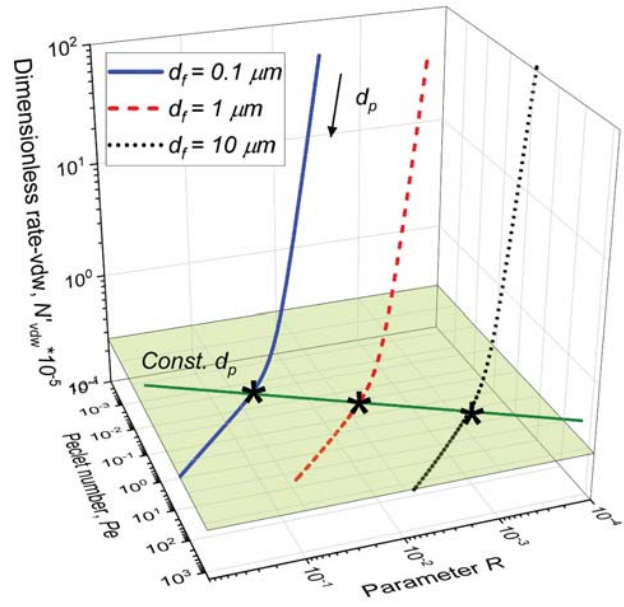

**Fig. S3.** The dimensionless rate of diffusion-vdw versus Pe and R.

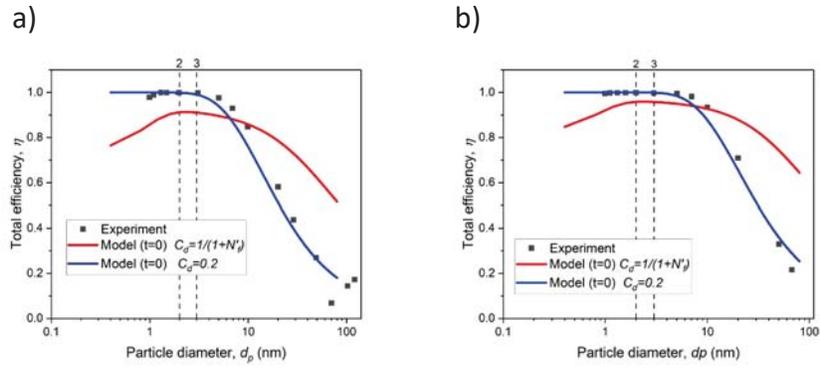

**Fig. S4.** The model results compared to the experiment (30) a) case 7, and b) case 8.

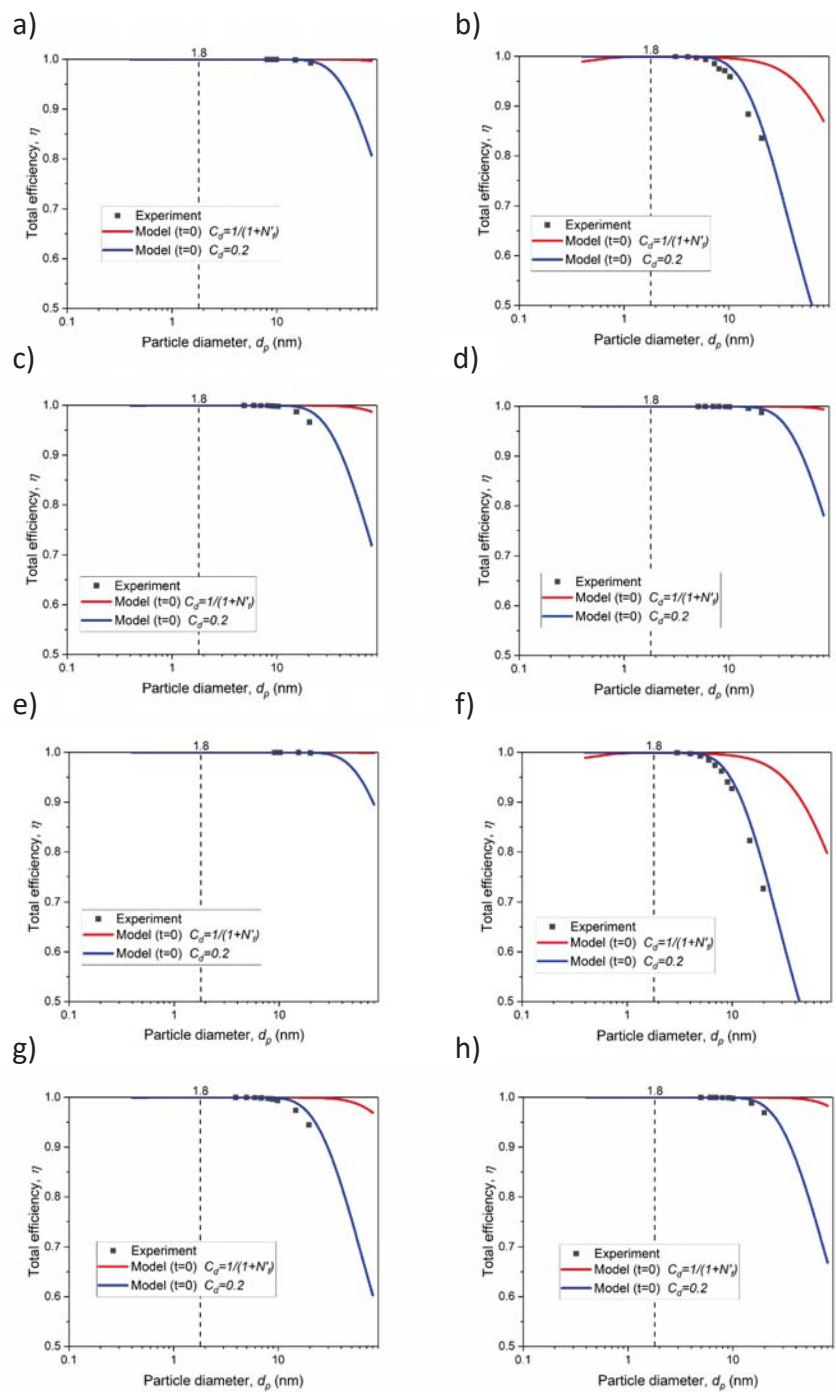

**Fig. S5.** The model compared to the experiment a) to h) are for cases 9 to 16, respectively.

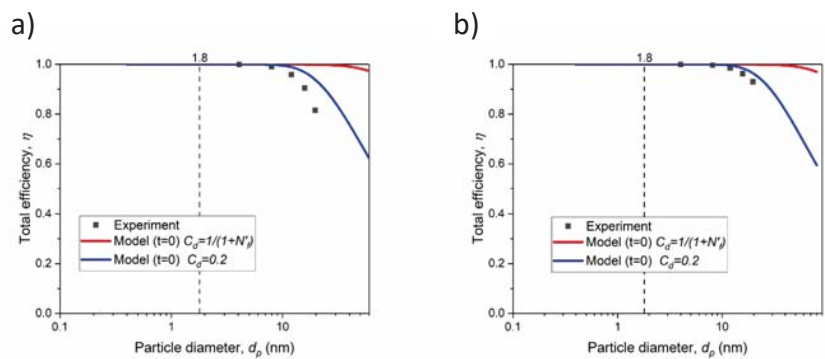

**Fig. S6.** The model compared to the experiment (32) a) to b) are for cases 17 and 18.

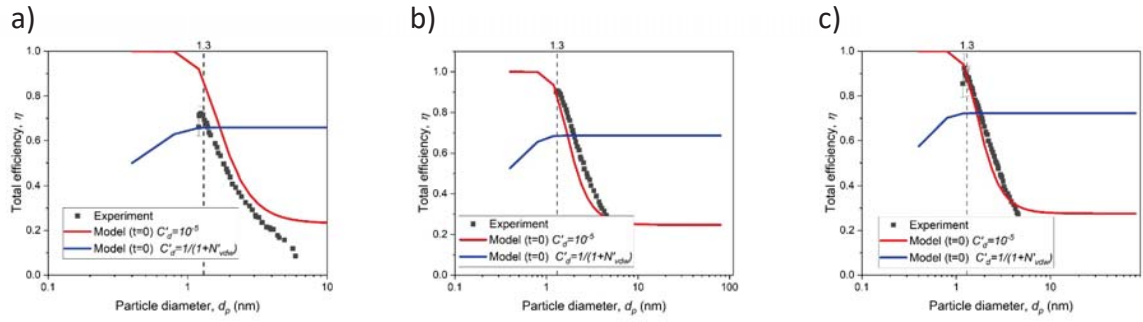

**Fig. S7.** The model compared to the experiment (34) a) case 1, b) case 2, and c) case 3.

**Table S1.** The properties of nanoparticles in the experimental data used for comparison

| No | Ref | Particle properties | | Filter properties | | | | | Operating parameters | |
|---|---|---|---|---|---|---|---|---|---|---|
| | | Particle material | Hamaker constant (j) | Fiber material | Fiber diameter, $d_f$ (m) | Filter thickness, L (m) | Solidity, $\alpha$ | Hamaker conctant (j) | Face velocity, u (m/s) | Temperature, T (K) |
| 1 | (34) | WOx | 2.12E-19 | stainless steel | 0.000102 | 0.000204 | 0.297 | 3E-19 | 0.15 | 293 |
| 2 | (34) | WOx | 2.12E-19 | stainless steel | 0.00005 | 0.0001 | 0.313 | 3E-19 | 0.15 | 293 |
| 3 | (34) | WOx | 2.12E-19 | Nickel | 0.000054 | 0.000108 | 0.335 | 4E-19 | 0.165 | 293 |
| 4 | (33) | Silver | 3.1E-19 | stainless steel | 0.00009 | 0.000203 | 0.3105 | 3E-19 | 0.0493 | 350 |
| 5 | (33) | Silver | 3.1E-19 | stainless steel | 0.00009 | 0.000812 | 0.3105 | 3E-19 | 0.0935 | 350 |
| 6 | (35) | NaCl | 7E-20 | stainless steel | 0.00004 | 0.00008 | 0.075 assumed | 3E-19 | 4.55E-05 | 293 |
| 7 | (36) | NaCl | 7E-20 | fiber glass | 1.18E-05 | 0.0004 | 0.0588 | 8.5E-20 | 0.025 | 293 |
| 8 | (36) | NaCl | 7E-20 | fiber glass | 9.1E-06 | 0.0004 | 0.0588 | 8.5E-20 | 0.025 | 293 |
| 9 | (32) | Silver | 3.1E-19 | fiber glass | 4.9E-06 | 0.00074 | 0.039 | 8.5E-20 | 0.053 | 293 |
| 10 | (32) | Silver | 3.1E-19 | fiber glass | 3.3E-06 | 0.00074 | 0.047 | 8.5E-20 | 0.053 | 293 |
| 11 | (32) | Silver | 3.1E-19 | fiber glass | 2.9E-06 | 0.00069 | 0.049 | 8.5E-20 | 0.053 | 293 |
| 12 | (32) | Silver | 3.1E-19 | fiber glass | 1.9E-06 | 0.00053 | 0.05 | 8.5E-20 | 0.053 | 293 |
| 13 | (32) | Silver | 3.1E-19 | fiber glass | 4.9E-06 | 0.00074 | 0.039 | 8.5E-20 | 0.1 | 293 |
| 14 | (32) | Silver | 3.1E-19 | fiber glass | 3.3E-06 | 0.00074 | 0.047 | 8.5E-20 | 0.1 | 293 |
| 15 | (32) | Silver | 3.1E-19 | fiber glass | 2.9E-06 | 0.00069 | 0.049 | 8.5E-20 | 0.1 | 293 |
| 16 | (32) | Silver | 3.1E-19 | fiber glass | 1.9E-06 | 0.00053 | 0.05 | 8.5E-20 | 0.1 | 293 |
| 17 | (37) | Silver | 3.1E-19 | fiber glass | 3.3E-06 | 0.00074 | 0.047 | 8.5E-20 | 0.15 | 293 |
| 18 | (37) | Silver | 3.1E-19 | fiber glass | 2.9E-06 | 0.00069 | 0.049 | 8.5E-20 | 0.15 | 293 |
| 19 | (38) | NaCl | 7E-20 | polyster | 0.000016 | 0.00045 | 0.118 | 6.5E-20 | 0.25 | 298 |
| 20 | (38) | NaCl | 7E-20 | polyster | 0.000016 | 0.0045 | 0.118 | 6.5E-20 | 0.145125 | 298 |
| 21 | (38) | NaCl | 7E-20 | polyster | 0.000016 | 0.0045 | 0.118 | 6.5E-20 | 0.072562 | 298 |
| 22 | (38) | NaCl | 7E-20 | polyster | 0.000016 | 0.0045 | 0.118 | 6.5E-20 | 0.034014 | 298 |
| 23 | (39) | WOx | 2.12E-19 | PVA | 1.83E-07 | 0.000012 | 0.0267 | 3.7E-19 | 0.069769 | 293 |

| 24 | (39) | WOx | 2.12E-19 | PVA | 1.64E-07 | 0.000008 | 0.0303 | 3.7E-19 | 0.070028 | 293 |
| --- | --- | --- | --- | --- | --- | --- | --- | --- | --- | --- |
| 25 | (39) | WOx | 2.12E-19 | PVA | 1.45E-07 | 0.000006 | 0.031 | 3.7E-19 | 0.070079 | 293 |
| 26 | (39) | WOx | 2.12E-19 | PVA | 1.62E-07 | 0.00001 | 0.0153 | 3.7E-19 | 0.068961 | 293 |
| 27 | (39) | WOx | 2.12E-19 | PVA | 1.63E-07 | 0.000014 | 0.0362 | 3.7E-19 | 0.070457 | 293 |
| 28 | (39) | WOx | 2.12E-19 | PVA | 1.52E-07 | 0.000004 | 0.0203 | 3.7E-19 | 0.069313 | 293 |
| 29 | (39) | WOx | 2.12E-19 | PVA | 1.22E-07 | 0.000017 | 0.0245 | 3.7E-19 | 0.069612 | 293 |
| 30 | (39) | WOx | 2.12E-19 | PVA | 1.47E-07 | 0.000029 | 0.0388 | 3.7E-19 | 0.070647 | 293 |

**Table S2.** The properties of intermediate nanoparticles using in experiments

| | Ref | Particle properties | | Filter properties | | | | | | Operating parameter | |
|---|---|---|---|---|---|---|---|---|---|---|---|
| | | Particle material | Hamaker constant (j) | Fiber material | Fiber diameter, $d_f$ (m) | Filter thickness*, L (m) | Solidity, $\alpha$ | Hamaker constant (j) | Filter diameter (m) | Face velocity, u (m/s) | Temperature, T (K) |
| 1 | (18) | WOx | 2.12E-19 | PVA | 1.87E-07 | 8.4E-06 | 0.0157 | 3.7E-19 | 0.025 | 0.137978 | 293 |
| 2 | (18) | WOx | 2.12E-19 | PVA | 1.82E-07 | 11.2E-06 | 0.0171 | 3.7E-19 | 0.025 | 0.138175 | 293 |
| 3 | (18) | WOx | 2.12E-19 | PVA | 1.91E-07 | 1.54E-05 | 0.009 | 3.7E-19 | 0.025 | 0.137184 | 293 |
| 4 | (18) | WOx | 2.12E-19 | PVA | 2.01E-07 | 8.4E-06 | 0.0144 | 3.7E-19 | 0.025 | 0.137796 | 293 |
| 5 | (18) | WOx | 2.12E-19 | PVA | 1.43E-07 | 7E-06 | 0.0283 | 3.7E-19 | 0.025 | 0.138344 | 293 |
| 6 | (18) | WOx | 2.12E-19 | PVA | 2.36E-07 | 1.54E-05 | 0.0107 | 3.7E-19 | 0.025 | 0.137323 | 293 |

\* The thickness of filters is measured using a digital micrometer by pressing the filter mat, thus the numbers are considered with 30% error.

**Table S3.** The properties of intermediate nanoparticles in experiments

| No | Ref | Particle properties ||| Filter properties ||||||| Operating parameter ||
|---|---|---|---|---|---|---|---|---|---|---|---|---|---|
| | | Particle material | Hamaker constant | Concentration (ppm) | Fiber material | Hamaker constant | Fiber diameter, $d_f$ | Solidity, $\alpha$ | Specific surface area | Weight (g) | Filter surface area (m$^2$) | Flow rate (LPM) | Temperature, T |
| 1 | (40) | VOC (toluene) | 5.4E-20 | 2000 | Carbon | 2.17E-19 | 4.6E-07 | 0.2 assumed | 1000 | 5 | π*0.028*0.1 | 0.5 | 323 |
| 2 | (40) | VOC (toluene) | 5.4E-20 | 2000 | Carbon | 2.17E-19 | 4.6E-07 | 0.2 assumed | 1500 | 5 | π*0.028*0.1 | 0.0010 | 323 |
| 3 | (41) | VOC (toluene) | 5.4E-20 | 50000 | Carbon | 2.17E-19 | 10–20 _m | 0.32 | 700–2500 | 6.2 | π*0.02*0.3 | 0.5 | 318 |
| 4 | (41) | VOC (toluene) | 5.4E-20 | 8750 | Carbon | 2.17E-19 | 10–20 _m | 0.32 | 700–2500 | 6.2 | π*0.02*0.3 | 0.5 | 318 |
| 5 | (41) | VOC (toluene) | 5.4E-20 | 4000 | Carbon | 2.17E-19 | 10–20 _m | 0.32 | 700–2500 | 6.2 | π*0.02*0.3 | 0.5 | 318 |

**References**

1. Li S, Marshall JS, Liu G, & Yao Q (2011) Adhesive particulate flow: The discrete-element method and its application in energy and environmental engineering. *Progress in Energy and Combustion Science* 37(6):633-668.
2. Kuwabara S (1959) The forces experienced by randomly distributed parallel circular cylinders or spheres in a viscous flow at small Reynolds numbers. *Journal of the physical society of Japan* 14(4):527-532.
3. Kirsh V (2000) The effect of van der Waals' forces on aerosol filtration with fibrous filters. *Colloid Journal* 62(6):714-720.
4. Mo Y, Turner KT, & Szlufarska I (2009) Friction laws at the nanoscale. *Nature* 457(7233):1116.
5. Cheng S, Luan B, & Robbins MO (2010) Contact and friction of nanoasperities: Effects of adsorbed monolayers. *Physical review E* 81(1):016102.
6. Johnson K, Kendall K, & Roberts A (1971) Surface energy and the contact of elastic solids. *Proceedings of the Royal Society of London A: Mathematical, Physical and Engineering Sciences*, (The Royal Society), pp 301-313.
7. Derjaguin BV, Muller VM, & Toporov YP (1975) Effect of contact deformations on the adhesion of particles. *Journal of Colloid and interface science* 53(2):314-326.
8. Maugis D (1992) Adhesion of spheres: the JKR-DMT transition using a Dugdale model. *Journal of colloid and interface science* 150(1):243-269.
9. Prokopovich P & Starov V (2011) Adhesion models: From single to multiple asperity contacts. *Advances in colloid and interface science* 168(1):210-222.
10. Hamaker HC (1937) The London—van der Waals attraction between spherical particles. *physica* 4(10):1058-1072.
11. Kim H-Y, Sofo JO, Velegol D, Cole MW, & Lucas AA (2007) Van der Waals dispersion forces between dielectric nanoclusters. *Langmuir* 23(4):1735-1740.
12. Israelachvili JN (2011) *Intermolecular and surface forces* (Academic press).
13. Dahneke B (1975) Kinetic theory of the escape of particles from surfaces. *Journal of Colloid and Interface Science* 50(1):89-107.
14. Peskir G (2003) On the diffusion coefficient: The Einstein relation and beyond.
15. Kisliuk P (1957) The sticking probabilities of gases chemisorbed on the surfaces of solids. *Journal of Physics and Chemistry of Solids* 3(1-2):95-101.
16. Roque-Malherbe RM (2018) *Adsorption and diffusion in nanoporous materials* (CRC press).
17. Nardin M & Papirer E (2006) *Powders and fibers: interfacial science and applications* (CRC Press).
18. Givehchi R, Li Q, & Tan Z (2018) Filtration of Sub-3.3 nm Tungsten Oxide Particles Using Nanofibrous Filters. *Materials* 11(8):1277.
19. Ortiz FG, Aguilera P, & Ollero P (2014) Modeling and simulation of the adsorption of biogas hydrogen sulfide on treated sewage–sludge. *Chemical Engineering Journal* 253:305-315.
20. Chauveau R, Grévillot G, Marsteau S, & Vallières C (2013) Values of the mass transfer coefficient of the linear driving force model for VOC adsorption on activated carbons. *Chemical engineering research and design* 91(5):955-962.

21. Patton A, Crittenden B, & Perera S (2004) Use of the linear driving force approximation to guide the design of monolithic adsorbents. *Chemical Engineering Research and Design* 82(8):999-1009.
22. Spurný K & Pich J (1963) Analytical methods for determination of aerosols with help of membrane ultrafilters. VI. On the mechanism of membrane ultrafilter action. *Collection of Czechoslovak Chemical Communications* 28(11):2886-2894.
23. Kangasluoma J, *et al.* (2013) Remarks on ion generation for CPC detection efficiency studies in sub-3-nm size range. *Aerosol Science and Technology* 47(5):556-563.
24. Wimmer D, *et al.* (2013) Performance of diethylene glycol-based particle counters in the sub-3 nm size range. *Atmospheric Measurement Techniques* 6(7):1793.
25. Kulmala M, *et al.* (2013) Direct observations of atmospheric aerosol nucleation. *Science* 339(6122):943-946.
26. Sem GJ (2002) Design and performance characteristics of three continuous-flow condensation particle counters: a summary. *Atmospheric research* 62(3-4):267-294.
27. Cai R, Jiang J, Mirme S, & Kangasluoma J (2018) Parameters governing the performance of electrical mobility spectrometers for measuring sub-3 nm particles. *Journal of Aerosol Science*.
28. Wiedensohler A (1988) An approximation of the bipolar charge distribution for particles in the submicron range. *J. Aerosol Sci.* 19(3):387-389.
29. Jiang J, *et al.* (2011) Transfer functions and penetrations of five differential mobility analyzers for sub-2 nm particle classification. *Aerosol science and technology* 45(4):480-492.
30. Kim CS, Bao L, Okuyama K, Shimada M, & Niinuma H (2006) Filtration efficiency of a fibrous filter for nanoparticles. *Journal of nanoparticle research* 8(2):215-221.
31. Payet S, Boulaud D, Madelaine G, & Renoux A (1992) Penetration and pressure drop of a HEPA filter during loading with submicron liquid particles. *Journal of Aerosol Science* 23(7):723-735.
32. Wang J, Chen D, & Pui D (2007) Modeling of filtration efficiency of nanoparticles in standard filter media. *Journal of Nanoparticle Research* 9(1):109-115.
33. Shin W, Mulholland G, Kim S, & Pui D (2008) Experimental study of filtration efficiency of nanoparticles below 20nm at elevated temperatures. *Journal of Aerosol Science* 39(6):488-499.
34. Heim M, Attoui M, & Kasper G (2010) The efficiency of diffusional particle collection onto wire grids in the mobility equivalent size range of 1.2–8nm. *Journal of Aerosol Science* 41(2):207-222.
35. Van Gulijk C, Bal E, & Schmidt-Ott A (2009) Experimental evidence of reduced sticking of nanoparticles on a metal grid. *Journal of Aerosol Science* 40(4):362-369.
36. Kim SC, Harrington MS, & Pui DY (2007) Experimental study of nanoparticles penetration through commercial filter media. *Journal of Nanoparticle Research* 9(1):117-125.
37. Rengasamy S, King WP, Eimer BC, & Shaffer RE (2008) Filtration performance of NIOSH-approved N95 and P100 filtering facepiece respirators against 4 to 30 nanometer-size nanoparticles. *Journal of occupational and environmental hygiene* 5(9):556-564.
38. Steffens J & Coury J (2007) Collection efficiency of fiber filters operating on the removal of nano-sized aerosol particles: I—Homogeneous fibers. *Separation and purification technology* 58(1):99-105.
39. Givehchi R, Li Q, & Tan Z (2016) Quality factors of PVA nanofibrous filters for airborne particles in the size range of 10–125 nm. *Fuel* 181:1273-1280.
40. Das D, Gaur V, & Verma N (2004) Removal of volatile organic compound by activated carbon fiber. *Carbon* 42(14):2949-2962.
41. Dwivedi P, Gaur V, Sharma A, & Verma N (2004) Comparative study of removal of volatile organic compounds by cryogenic condensation and adsorption by activated carbon fiber. *Separation and Purification Technology* 39(1-2):23-37.